\documentclass{amsart}

\usepackage[utf8]{inputenc}
\usepackage{color}
\usepackage{enumitem}
\usepackage{amsaddr}
\usepackage{hyperref}

\usepackage{graphicx}
\usepackage{natbib}
\newcommand{\E}{\mathsf{E}}
\usepackage{xcolor}
\usepackage{algorithmic}
\usepackage{algorithm}

\usepackage{bm}

\newcommand{\Un}{\ensuremath{\mathsf{U}}}
\newcommand{\Var}{\ensuremath{\mathrm{Var}}}

\theoremstyle{remark}
\newtheorem*{remark}{Remark}
\newcommand{\FB}{F^{(B)}}

\begin{document}

\title[Multivariate permutation tests]{Multivariate quantile-based permutation tests with application to functional data}


\author{Zdeněk Hlávka, Daniel Hlubinka, Šárka Hudecová}
\address{Charles University, Faculty of Mathematics and Physics,\\ Department of Statistics, Prague, Czech Republic}
\email{hlubinka@karlin.mff.cuni.cz}


\date{\today}

\maketitle

\begin{abstract}
Permutation tests enable testing statistical hypotheses 
in situations when the  distribution of the test
statistic is complicated or not 
available. In some situations, the test statistic under investigation is multivariate, with the multiple testing problem being an important example. The corresponding
multivariate permutation tests are then typically  based on a suitable
  one-dimensional transformation of 
  the vector of partial permutation p-values via so called combining functions. 
  This paper proposes a new approach 
  that utilizes the optimal measure transportation concept. The final single p-value is computed from the empirical  center-outward distribution function of the permuted multivariate test statistics. 
  This method avoids computation of the partial p-values and it
  is easy to be implemented. In addition, it allows to compute and interpret contributions of the components of the multivariate test statistic to the non-conformity score and to the rejection of the null hypothesis. Apart from this method, the measure transportation is applied also to the vector of partial p-values as an alternative to the classical combining functions. Both techniques are compared with the standard approaches using various practical examples in a Monte Carlo study. An application on a functional data set is provided as well.
\end{abstract}

\section{Introduction}

Permutation tests provide a powerful tool for testing statistical hypotheses 
in situations when the exact or asymptotic distribution of the test
statistic  is not readily
available. Their main advantage is that they  
require minimal assumptions about the underlying data distribution.
This is particularly useful 
in functional data analysis (FDA), where
the data consist of functions over a continuous time domain and a parametric model would be too restrictive. Permutation tests have been successfully applied in functional regression \citep{cardot2007, baoyi2021}, functional two-sample problem and ANOVA \citep{kashlak2022, hlavka2022functional}, in testing the equality of distribution functionals \citep{bugni2021}, and in many other areas of FDA.


The application of permutation
tests is straightforward for a univariate test statistic, where the significance 
is evaluated using so called permutation distribution obtained by random permutations of the data. The corresponding p-value is computed as a proportion of permuted test statistics which are more extreme than the value computed from the original data. 
If the test statistic is multivariate then this approach cannot be used directly due to 
the lack of
linear ordering in $\mathbb{R}^d$ for $d\geq 2$, because then
 one cannot 
easily decide whether the observed value of the test statistic is extreme enough
to reject the null hypothesis. 
Such situation arises in various practical setups, particularly within the multiple testing problem where the multiple test statistics can be seen as components of one multivariate test statistic.
 In that situation the Bonferroni
inequality may be used to control the level of the test but it
is well-known to be quite conservative, so various modifications have been proposed.
A common approach is to compute so called partial $p$-values for each component of the
vector test statistic 
and then combine them into a final single p-value  
used for the final decision about the hypothesis \citep{kost,vovk,zhang23}.
Within the framework of permutation tests, nonparametric
combining functions allow to transform several dependent partial $p$-values,
obtained by the univariate permutation tests applied on  each
component of the vector test statistic, 
into a single p-value, with  
 the Liptak, Tippett, and Fisher
combining functions being the most popular ones \citep[Chapter~6]{pesarin2001multivariate}.

In this paper, we introduce a new method for dealing with multivariate permutation tests. Namely, we propose 
to utilize the concept of measure transportation and the corresponding multivariate quantiles from
\cite{hallin_2021} that allows to define 
a permutation p-value for a
multivariate test statistic directly, without the need to compute the partial p-values and then  combining them. 
The main advantages of the proposed approach  are: 
\begin{itemize}
\item it is fully distribution-free and only mild assumptions about exchangeability under the null hypothesis are required,
\item there is no need to compute the partial permutation $p$-values and to 
choose a combining function 
since the final single permutation $p$-value is computed directly from the \emph{multivariate} test statistic and its permutation counterparts, 
\item the approach allows to calculate and interpret
contributions of the components of the vector test statistic to the rejection of the null hypothesis.
\end{itemize}
Furthermore, the method can be used for a vector test statistic even in situations when its components do not naturally correspond to some partial null hypotheses and if its components are both one-sided and two-sided.

Apart from this approach, we also consider a setup where the measure transportation is applied to the vector of the partial p-values, so the transport in some sense substitutes the combining functions. This method might be attractive to
researchers who are used to working with partial p-values, or when only partial p-values are available, as, e.g., in clinical meta-analyses. However, our Monte Carlo study suggests that a transport of the original test statistic typically leads to a more powerful test.


A slightly related  
situation arises in the problem of synthesizing inference from multiple data sources or studies, also known as fusion learning. Recently, a depth confidence distribution combining statistical data depth function and confidence distribution was proposed by \cite{liu2022}  as a nonparametric tool for the fusion learning that overcomes some of the shortcomings and limitations of existing methods.
The notation of data depth is  related to the measure transportation, but 
the latter one allows to interpret not only the depth of a point  but also its direction from the center, which shows to be useful in the multiple testing problem.




The paper is organized as follows. In Section~\ref{sec:2} we review
multivariate tests statistics and the classical construction of scalar 
permutation $p$-values. We also provide several illustrations for the considered setup, with an emphasis on an application on functional data in Section~\ref{sec:2.2}. 
 Section~\ref{sec:3} describes a construction of multivariate quantiles based on the optimal measure transport to a spherically uniform distribution.  The construction of permutation p-values based on this approach is discussed in Section~\ref{sec:4}.  Section~\ref{sec:sim} contains a Monte Carlo study for various examples, while Section~\ref{sec:fanova} illustrates the approach to functional ANOVA. 

\section{Classical multivariate permutation tests}\label{sec:2}

A permutation test is a statistical method used to assess the significance of an observed sample statistic by comparing it to a distribution of possible values obtained by permuting the observations from the sample. 
Consider a null hypothesis $H_0$ and the corresponding scalar test statistic $T$. Assume that the data are exchangeable under $H_0$ and let $T_0$ stand for the value of the test statistic computed from the observed data. 
For a chosen $B$ repetitions,
one randomly shuffles the labels of the observations and
calculates the test statistics for the permuted data sets to obtain $T_1,\dots,T_B$.
The null hypothesis 
is rejected whenever the observed $T_0$ 
fails to concur with 
$T_1,\dots,T_B$, and 
 the p-value is calculated as the proportion of permuted test statistics that are more extreme than the observed test statistic.


In various situations, the test of $H_0$ is based on a multivariate test statistic $\bm T$. Then one can still use the permutation principle which now leads to multivariate permutation test statistics $\bm{T}_1,\dots,\bm{T}_B$. The classical approach is then to compute componentwise partial p-values. 
The resulting vector of partial p-values then needs to be combined into a single final p-value $p_0$, and 
the value $(1-p_0)^2$ can  be interpreted as a multivariate nonconformity score because it measures, in some sense, the deviation of the data from the null hypothesis.

We illustrate this approach with a well-known toy example considering three independent
random samples from shifted continuous distributions. Namely, let $f$ be a density of an absolutely continuous distribution with a finite variance, and 
$X_{1,1},\dots,X_{1,n_1}$, $X_{2,1},\dots,X_{2,n_2}$,
and $X_{3,1},\dots,X_{3,n_3}$ be random  samples with density $f(x-\mu_1)$, $f(x-\mu_2)$, and $f(x-\mu_3)$, respectively. 
Consider  the null hypothesis
\begin{equation}\label{eq:H0}
H_{0}: \mu_1 = \mu_2 =\mu_3
\end{equation}
against a general alternative, i.e., a three-sample problem. Note that $H_{0}$ is equivalent to the pair of hypotheses:
\begin{eqnarray*}
&H_{0,1}:& \mu_1 = \mu_2\\
&H_{0,2}:& \mu_1 = \mu_3.
\end{eqnarray*}
Each hypothesis $H_{0,j}$, $j=1,2$, may be tested separately by a two-sample t-test. 
Assuming normality, the corresponding test statistics $T_{1}$ and $T_{2}$ follow $t$
distribution under the null hypothesis. However, $T_{1}$ and $T_{2}$
are not independent and, despite knowing their marginal distributions,
their bivariate joint distribution is not tractable although the
F-statistic can be used to test the null hypothesis $H_{0}$.

 The normality assumption is not needed if we rely on asymptotics or use 
permutation
tests. Focusing on the latter approach, 
the partial permutation p-values can be obtained as
follows: 

\begin{enumerate}[label=(P.{\arabic*})]
\item Calculate the two-sample t-test statistics $T_{1,0}$ and $T_{2,0}$ from the original sample.
\item Consider a random permutation $\pi$ of the set
  $\{1,\dots,n\}$. Let $n=n_1+n_2+n_3$, set
  $Z_1=X_{1,1}, Z_2=X_{1,2},\dots, Z_{n} = X_{3,n_3}$, and define
  $X^*_{1,1} = Z_{\pi(1)}, \dots, X^*_{1,n_1} = Z_{\pi(n_1)},
  X^*_{2,1} = Z_{\pi(n_1+1)},\dots,X^*_{3,n_3} = Z_{\pi(n)}$ so that
  $X^{*}_{j,i}$ is a random permutation sample of the original
  (pooled) sample.
  Calculate the values of the test statistics $T_{1}$ and
  $T_{2}$ from the permuted sample. 
  \label{p2}
\item Carry out $B$ independent repetition of the step~\ref{p2} so
  that the bivariate sequence
  $\{(T_{1,b},T_{2,b}), b=1,\dots,B\}$ forms a random sample
  from the permutation distribution of $(T_{1,0},T_{2,0})$.
\item Compute partial p-value for the  null hypothesis $H_{0,j}$ 
as\[
  p_{j}=(B+1)^{-1} \Bigl\{1+\sum_{b=1}^B
  \boldsymbol{1}(|T_{j,b}|\geq |T_{j,0}|)\Bigr\}.
  \]
\end{enumerate}

  In this example, we have reformulated the null hypothesis $H_0$ in \eqref{eq:H0} in terms of two pairwise comparisons for couples $(\mu_1,\mu_2)$ and $(\mu_1,\mu_3)$. In that case, the permutation procedure leads to a two-dimensional vector of p values $\bm{p}=(p_1,p_2)^\top$ composed of two partial p-values.
 An alternative approach is to consider all pairwise comparisons, which would lead, in this example, to a three-dimensional test statistic $\bm{T}$. 
 In a general situation, one can have a $d$-dimensional test statistic $\bm{T}$. The permutation procedure then proceeds in an analogous way, leading to a $d$-dimensional vector of p-values $\bm{p}$,  which needs to be combined to a single final p-value used for evaluation of the null hypothesis.
 This step is often based on combining functions, as  described in the  next section.

\subsection{Combining functions} \label{sec:21}


Let $\bm p=(p_1,\dots,p_d)^\top$ be a $d$-dimensional vector of partial p-values obtained by a permutation procedure performed on the components of $\bm T$. Let $f:[0,1]^d\to[0,1]$ be a selected combining function. 
 The multivariate permutation test
 of $H_{0}$ based on a combining function $f$  proceeds in the following way:
\begin{enumerate}[label=(C.{\arabic*})]
\item For $b=1\dots,B$ and $j=1,\dots,d$ define 
\[
p_{j,b}=(B+1)^{-1} \left\{1+\sum_{s=1}^B \boldsymbol{1}(|T_{j,s}|\geq
|T_{j,b}|)\right\}.
\]
a permutation p-value which would correspond to the test statistic $T_{j,b}$. Let $\bm{p}_b = (p_{1,b},\dots,p_{d,b})^\top$.
\item 
Define 
 $Q_{0}=f(\bm p)$
  and $Q_{b}=f(\bm p_b)$, for
  $b=1,\dots,B$.
\item The multivariate permutation p-value is then defined as
  \[
  p_0=
  (B+1)^{-1} \Bigl\{1+\sum_{b=1}^B
  \boldsymbol{1}(Q_{b}\geq Q_{0})\Bigr\}.
  \]
\end{enumerate}

Commonly used combining functions are the Tippett combining
function $f_T(\bm{p})=\max_{1\leq j\leq d}(1-p_{j})$, the Liptak combining function
$f_L(\bm{p}) = \sum_{j=1}^d \Phi^{-1}(1-p_{j})$, and the Fisher omnibus
combining function $f_F(\bm{p})=-2\sum_{j=1}^d\log(p_{j})$.

\begin{figure}
\includegraphics[width=\textwidth]{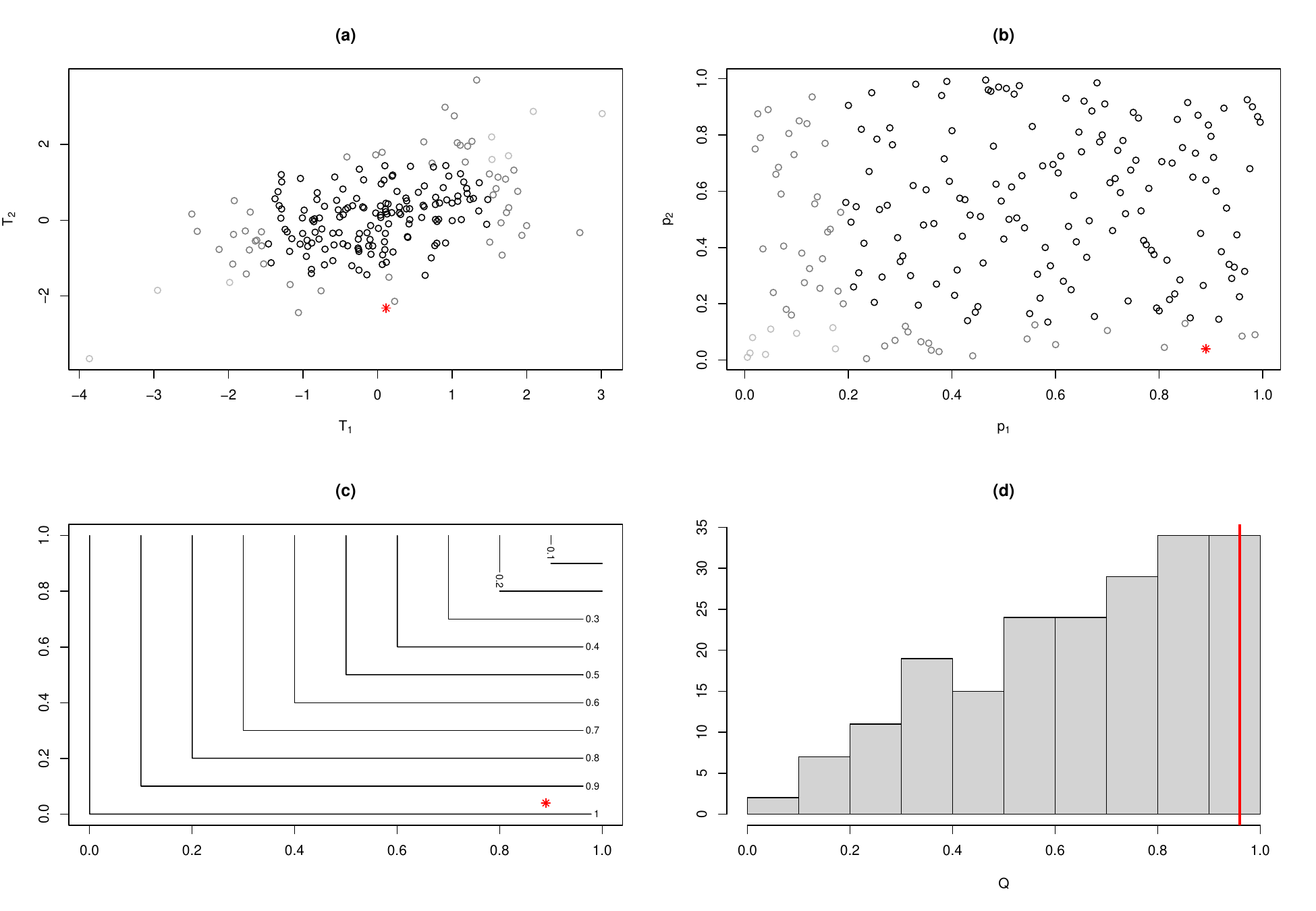}
\caption{Bivariate permutation test with Tippett combining function: (a) bivariate
permutation test statistics $(T_{1,b},T_{2,b})$, red star
denotes $(T_{1},T_{2})$, $B=199$, (b) partial permutation p-values 
$(p_{1,b},p_{2,b})$, red star denotes the observed 
$(p_{1},p_{2})$, (c) Tippett combining function $f_T(.,.)$, 
(d) histogram of $Q_{b}$, vertical red line denotes $Q_{0}$, 
multivariate permutation p-value $p_0=0.065$.}
\label{pokus1-papr}
\end{figure}

A simulated example using the Tippett combining function with $n_1=n_2=n_3=8$, 
$\mu_1=\mu_2=1$, and $\mu_3=2$ with $\sigma=1$ is plotted in
Figure~\ref{pokus1-papr}. Following the multivariate permutation test algorithm, with
each step illustrated in Figure~\ref{pokus1-papr}, we
arrive to the multivariate permutation p-value $p_0=0.065$ 
and we do not reject the 
null hypothesis $H_{0}$ on confidence level $\alpha=0.05$. 
This result 
strongly depends on the choice of the combining function  and different p-values could be obtained simply by using another combining function.
In our simulated example, p-values 0.140 and 0.360 are obtained, respectively, by applying the Fisher and Liptak combining function.

\begin{remark}
Another possibility for obtaining 
a final single p-value is via a suitable scalar function of 
$\bm{T}_0,\bm{T_1},\dots,\bm{T}_B$, 
for example, their $l_p$ norm for some $p\in[1,\infty]$.
This approach is called a \emph{direct combination} but it is recommended only for test statistics with the same marginal asymptotic distributions and for a large sample size, see \citet[Section 6.2.4 (h)]{pesarin2001multivariate}. 
\end{remark}

\subsection{More general setups} \label{sec:2.2}

 The simple three samples problem from the beginning of this section serves mainly for an illustration of the multivariate permutation testing approach, but the problem of multiple testing occurs, of course, in many more complex situations.   

\subsubsection{Testing equality of $r$ moments.}\label{sec:2.3.1}
A multivariate test statistic naturally arises when one wants to test equality of moments up to a prescribed order $r\geq 2$ for  $K\geq 2$ samples. Assume that $X_{j,1},\dots,X_{j,n_j}$ are independent random variables from a  distribution $F_j$ with 
$\int_{-\infty}^\infty |x|^r \mathrm{d} F_j(x)<\infty$ for  $j=1,\dots,K$ and these $K$ samples are independent.
Denote as  $\mu_s^{(j)}=\int_{-\infty}^{\infty} x^s \mathrm{d} F_j(x)<\infty$, $s=1,\dots,r$ the non-central moment for $j$-th sample.  Consider the joint null hypothesis 
\[
H_0: \mu_s^{(1)}=\dots=\mu_s^{(K)} \quad  \text{for all }s=1,\dots,r,
\]
 which can be decomposed into $r$ marginal hypotheses $H_{0,s}:  \mu_s^{(1)}=\dots=\mu_s^{(K)}$ for $s=1,\dots,r$. Each of them can be tested in terms of a suitable version of ANOVA  applied on $X_{i,j}^s$, or alternatively further decomposed  to pairwise comparisons, leading generally to a $d$ dimensional test statistic $\bm{T}$. If the data are exchangeable under the null hypothesis $H_0$, then the permutation principals may be used for $\bm{T}$.

\subsubsection{Functional data}\label{sec:func}
There are many real-world situations where the system is observed over a continuous time domain,  such as economics, meteorology, medicine, and more. Functional Data Analysis (FDA) is an active area of statistics, which deals with data in the form of functions. Permutation techniques, analogous to those described in Section~\ref{sec:21},  serve as a standard tool in FDA testing, because the (asymptotic) distribution of a considered test statistic is typically very complex even under the null hypothesis. 

 In what follows we assume that a functional random variable $X$ is a random element in a Banach space  $\mathcal{X}=\mathcal{L}_2[0,1]$ of  square integrable functions on $[0,1]$. The space $\mathcal{X}$ is equipped with an inner product
$\langle u,v\rangle = \int_{0}^1 u(t)v(t)\mathrm{d} t$, $u,v\in\mathcal{X}$, and induced norm $\|u\| = \sqrt{\langle u,u\rangle}$.
 Namely, if $(\Omega,\mathcal{A},\mathsf{P})$ is a probability space, then we say that $X:\Omega\to\mathcal{X}$ is a functional random variable if  the real-valued random variable $\|X\|:\Omega\to\mathbb{R}$ has finite second moment, i.e., if 
$\mathsf{E} \int_{0}^1 X^2(t)\mathrm{d} t<\infty$.   A distribution of a functional random variable $X$ can be described by a characteristic function, which is defined  as
\[
\phi(w) = \E \mathrm{e}^{i\langle w,X\rangle}, \quad w\in\mathcal{X}. 
\]
Other useful characteristics are the mean function $\mu(t)=\E X(t)$ and the covariance operator $C:\mathcal{X}\to\mathcal{X}$ defined as $C(y)=\E\langle X-\mu,y\rangle(X-\mu)$, $y\in\mathcal{X}$.

Consider a functional $K$-sample problem. Let $X_{j,1},\dots,X_{j,n_j}$ be independent and identically distributed functional observations  with a characteristic function $\phi_j$, mean $\mu_j$ and covariance operator $C_j$ for $i=1,\dots,n_j$, $j=1,\dots,K$, and let the $K$ samples be independent.
A general null hypothesis of equality of the whole distributions  is stated as  
\[
H_0:\phi_1=\dots=\phi_K. 
\] 
For the two sample problem, $K=2$, \citet{hlavka2022functional} propose a test 
statistic $T_{12}$ for testing $H_0$ based on empirical analogous of the
characteristic functions $\phi_j$, $j=1,2$. The null hypothesis is rejected for large
values of $T_{12}$ (one-sided test).  The significance of $T_{12}$ is evaluated via a
permutation test, which can be applied here, because all the observations are
exchangeable under $H_0$, and which is a complete analogy to the procedure  described
in Section~\ref{sec:21}.  For $K>2$, the null hypothesis $H_0$ can be reformulated in
terms of all $\genfrac(){0pt}{1}{K}{2}$  pairwise comparisons $H_{0,kl}: \phi_k=\phi_l$ for all
$1\leq k<l\leq K$. Each $H_{0,kl}$ can be tested using the permutation two-sample test
statistic $T_{k,l}$, leading to a $d=\genfrac(){0pt}{1}{K}{2}$ dimensional vector of all test
statistics $\bm{T} $ and $d$ dimensional vector of partial permutation p-values
$\bm{p}$. 

Alternatively, one may focus on  the equality of the mean functions in terms of the null hypothesis
\[
H^M_0: \mu_1=\dots=\mu_K,
\]
 or equality of covariance operators
\[
H^C_0: C_1=\dots=C_K.
\]  
Various tests have been proposed for testing the hypothesis $H_0^M$ or $H_0^C$ for a general $K$ sample problem \citep{zhang2013analysis, cuevas04, gorecki15,cuesta,gorecki19}. However, one may want to test these two hypothesis jointly, i.e. to consider $H_0^{CM}:  \mu_1=\dots=\mu_K \ \& \ C_1=\dots=C_K$. 
If the data are exchangeable under the null hypothesis $H_0^{CM}$, then one can use permutation tests for testing $H_0^M$ and $H_0^C$ separately, leading to a two-dimensional test statistic $\bm{T}$ and a vector of partial permutation p-values $\bm{p}$. 
 This approach is illustrated on a real data set in Section~\ref{sec:fanova}.

 \subsection{Two-sided vs one-sided test statistics} \label{sec:2.3}

 In the following we will sometimes need to distinguish whether the test statistic $\bm{T}$ under investigation is two-sided, one-sided or a combination of both. The meaning of these terms is explained in the next paragraphs.

The multivariate permutation test from 
Section~\ref{sec:21} is based on a two-dimensional test statistic $\bm{T}=(T_1,T_2)^\top$, whose components correspond to the null hypotheses $H_{0,j}$, $j=1,2$. Each of the two partial tests is two-sided in a sense that the null hypothesis $H_{0,j}$ is  rejected for large values of $|T_j|$. 
However, there are various statistical tests 
which are one-sided, so
 the null hypothesis is
rejected only when the corresponding univariate test statistic is larger than
a given critical value while it is not rejected for
 small values (e.g. $\chi^2$ test, $F$ test, one sided $t$-test etc.). A similar situation can occur also with a multivariate test statistic. For instance in some specific regression problems one may want to test the joint significance of the vector of regression parameters against an alternative that all the effects are positive, i.e. to test $H_0:\bm{\beta}=\bm{0}$ against $H_1:\bm{\beta}>\bm{0}$, where the inequality is taken componentwise. Then the least squares estimator $\widehat{\bm{\beta}}$ can be considered as a multivariate test statistic and the corresponding testing problem is one-sided. Some other examples of one-sided multivariate tests are presented in Section~\ref{sec:func}. 

 Finally, some practical situations can lead to a multivariate test statistic $\bm{T}$ composed by two-sided and one-sided test statistics. For instance, in a two-sample setup, one may want to test the joint hypothesis of equality of means and variance while considering a two-sided alternative for the means and one-sided alternative for the variances.


\section{Multivariate quantiles based on measure transportation}\label{sec:3}

There exist several approaches intending to define multivariate
ranks and quantiles, often based on some depth function, directions, or optimal ellipsoids; see, e.g., \citet{chaudhuri1996, pokorny2022,
  chandler2022, hlubinka2022,hlubinka2015, koshevoy1998}.
  Recently, results from the optimal measure transportation \citep{villani2021, figalli2021} have been successfully applied in this context, resulting in the so called center-outward (CO) distribution function and quantiles  \citep{galichon2017, hallin_2021}. The corresponding ranks and signs have been applied in various multivariate statistical  problems, see, e.g., \citet{shi, marcVAR,hallin_2022}. In what follows we utilize this technique since it 
offers possibility to construct also multivariate
versions of ``one-sided'' tests.


Let  $\Un_d$ denote a distribution on the unit ball
$\mathbb{S}_d = \{\bm{x} \in \mathbb{R}^d; \|x\| \leq 1\}$ corresponding to a random variable $U \bm{S}$, 
where $U$ is  uniformly distributed on $[0,1]$ and it is independent of $\bm{S}$, which has a uniform distribution
on the unit sphere $\mathcal{S}_{d-1} = \{\bm{x} \in \mathbb{R}^d; \|x\| = 1\}$.



Let $P$ be an absolute continuous
distribution  on $\mathbb{R}^d$. \citet{McCann1995} proved that
there exists a convex map $\psi_P$ such that its gradient
$\nabla\psi_P$ pushes $P$ to $\Un_d$, i.e., if a random vector $\bm{Z}\sim P$, the random vector $\nabla\psi_P(\bm{Z}) \sim \Un_d$. Moreover, the gradient
$\nabla\psi_P$ is $P$-a.s. unique. If the distribution $P$ has finite
second moments, the transformation $\nabla\psi_P$ is optimal with
respect to the quadratic loss function, i.e.,
$\mathbb{E}\bigl\|\bm{Z} -\nabla\psi_P(\bm{Z})\bigr\|^2=\min_{\bm{Y}\sim \Un_d} \mathbb{E}\bigl\|\bm{Z} -\bm{Y}\bigr\|^2$, 
see \citet[Chapter~2]{figalli2021}. The mapping $\nabla\psi_P$ is called the
\emph{center-outward distribution function} and it is denoted as $F_\pm$,
see \citet{hallin_2021}.
Since $F_{\pm}(\bm{Z})\sim U_d$  is spherically symmetric, and $\|F_{\pm}(\bm{Z})\|$ is uniformly distributed on $[0,1]$,
it is natural to define 
the
$\alpha$-quantile contour of $P$ as the set
$\mathcal{T}_{P,\alpha} = \{\bm{x}; \|F_\pm(\bm{x})\| = \alpha\}$, and
the $\alpha$-quantile region of $P$ as
$\mathbb{T}_{P,\alpha} = \{\bm{x}; \|F_\pm(\bm{x})\| \leq \alpha \}$.

Consider a random sample $\bm{Z}_1,\dots,\bm{Z}_n$ from 
the 
distribution $P$. 
The empirical counterpart $F_{\pm}^{(n)}$ of $F_{\pm}$ is defined as the mapping
 from $\bm{Z}_1,\dots,\bm{Z}_n$  to a given regular grid $\mathcal{G}_n$ of $n$ points in the unit ball $\mathbb{S}_d$ such that the mapping minimizes the sum of squared Euclidean distances $\sum_{i=1}^n \|F_{\pm}^{(n)}(\bm{Z}_i) - \bm{Z}_i\|^2$. 
A Glivenko-Cantelli result holds for $F_{\pm}^{(n)}$ and $F_{\pm}$ whenever  $F_{\pm}^{(n)}$ is computed on a sequence of grids $\{\mathcal{G}_n\}$ such that the discrete uniform distribution on $\mathcal{G}_n$ converges weakly to $\Un_d$. 

Using $F_{\pm}^{(n)}$ 
the empirical
$\alpha$-quantile region can be defined as a collection of observed points
\[
\{\bm{Z}_i: \| F_{\pm}^{(n)}(\bm Z_i)\|\leq \alpha\}.
\]
The
problem of turning such set into continuous contours enclosing compact regions is studied in \citet[Section 3]{hallin_2021}.

\subsection{Transportation grid}\label{sec:grids}
The computation of the sample CO distribution function requires a choice of the target grid $\mathcal{G}_n$. \citet{hallin_2021} recommends to use a product form grid, but some other types of grids also proved   to be useful in the statistical testing, see, e.g., \citet{Mordant23} or \citet{hudecova23}. 

The construction of a grid $\mathcal{G}_n$ in $\mathbb{S}_d$, $d\geq 2$, often makes use of so called low-discrepancy points
from $[0,1]^r$ for  $r=d$ or $r=d-1$. 
Recall that a low-discrepancy sequence $\{\bm{x}_k\}_{k=1}^K$ of $K$ points in  $[0,1]^r$ is a deterministic sequence designed to be ``as uniform as possible'' 
in a sense that it  minimizes the discrepancy between the distribution of $\{\bm{x}_k\}_{k=1}^K$ and a uniform distribution in  $[0,1]^r$, see \cite[Section 1.2]{fang}. 
Roughly speaking, the proportion of points from $\{\bm{x}_k\}_{k=1}^K$  falling into any rectangle $A\subset[0,1]^r$ should be close to the volume of $A$. 
There are various approaches to the construction of such sequences, the most popular ones are   
Halton sequences \citep{halton},  Sobol sequences \citep{sobol}, or 
good-lattice-point sets (GLP) \citep[Section 1.3]{fang}.

In this paper, we consider the following two types of grids:
\begin{itemize}
\item A product form grid (abbreviated as $\mathcal{P}$) 
as a union of $n_0$ replicas of $\bm 0$ and points
\begin{equation*} 
{\bm g}_{ij}=\frac{i}{n_R+1} \bm{s}_j, \quad i=1,\dots,n_R, \ j=1,\dots,n_S,
\end{equation*}
for $n=n_R\cdot n_S+n_0$, where $\bm{s}_j$ are unit vectors from  the unit sphere $\mathcal{S}_{d-1}$ 
obtained as a suitable transformation (see below) of low-discrepancy points in  $[0,1]^{d-1}$. Moreover, $n_0$ is negligible compared to $n_R$ and $n_S$, typically $n_0\in\{0,1\}$.
\item A non-product form grid (abbreviated as $\mathcal{N}$)  $\mathcal{G}_n=\{\bm{g}_i\}_{i=1}^n$ 
obtained from a low-discrepancy sequence of   points $\bm{x}_1,\dots,\bm{x}_n$ in $[0,1]^d$ such that
\[
\bm{g}_i = x_{i,1} \cdot \bm s_i, \quad i=1,\dots,n,
\]
where $\bm{x}_i=(x_{i,1},\dots,x_{i,d})^\top$  and the vector $(x_{i,2},\dots,x_{i,d})^\top \in[0,1]^{d-1}$ is transformed to a vector $\bm{s}_i\in\mathcal{S}_{d-1}$ using a suitable transformation (see below).
\end{itemize}

The transformation $\tau:[0,1]^{d-1}\to\mathcal{S}_{d-1}$ of points from $[0,1]^{d-1}$ to the vectors from the unit sphere has to be such that if a random vector $\bm{X}$ has a uniform distribution on $[0,1]^{d-1}$ then $\bm{S}=\tau(\bm{X})$ has a uniform distribution on $\mathcal{S}_{d-1}$. Such $\tau$ is described in \citet[Section 1.5.3]{fang} for general $d\geq 2$. Particularly, if $d=2$ then  the corresponding transformation $\tau:[0,1]\to\mathcal{S}_1$ is simply
\[
\tau(x)=\bigl(\cos(2\pi x), \sin(2\pi x)\bigr)^\top.
\]
In addition, for $d=2$ the sequence of $K$ points 
$\{(2k - 1)/{(2K)}\}_{k = 1}^K$ has the lowest possible discrepancy, so the directions $\bm{s}_j$ of the grid $\mathcal{P}$ can be chosen completely regularly for any $n_S$ using this sequence and the transformation $\tau$. 
For $d=3$ the suitable transformation $\tau:[0,1]^2\to \mathcal{S}_{2}$ is
\[
\tau(\bm x) = \Bigl(1-2 x_1,2 \sqrt{x_1(1-x_1)}\cos(2\pi x_2), 2\sqrt{x_1(1-x_1)}\sin(2\pi x_2)  \Bigr)^\top
\]
for $\bm x= (x_1,x_2)^\top$.

\begin{figure}[htb]
\includegraphics[width=10cm]{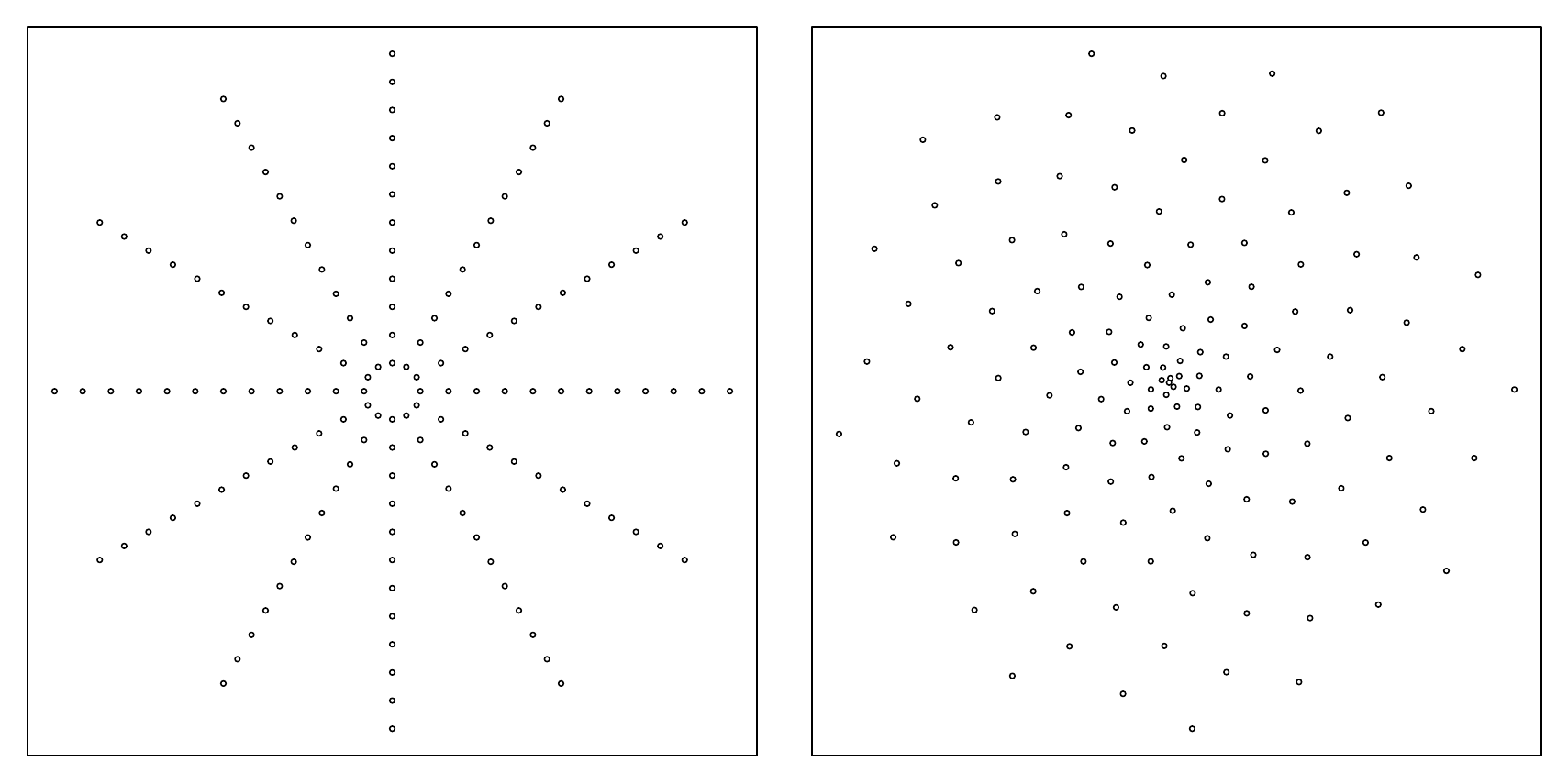}
\caption{ Two different grids of $n=144$ points in $\mathbb{S}_2$: 
product-form grid  $\mathcal{P}$ with $n_R=n_S=12$ and $n_0=0$ (left
panel) and a non-product grid  $\mathcal{N}$ based on a GLP set
generated by the vector $(1,89)^\top$ (right panel).}\label{pokus1-body}
\end{figure}

\begin{figure}[htb]
  \includegraphics[trim=3cm 3cm 3cm 3cm,clip=true,width=6.2cm]{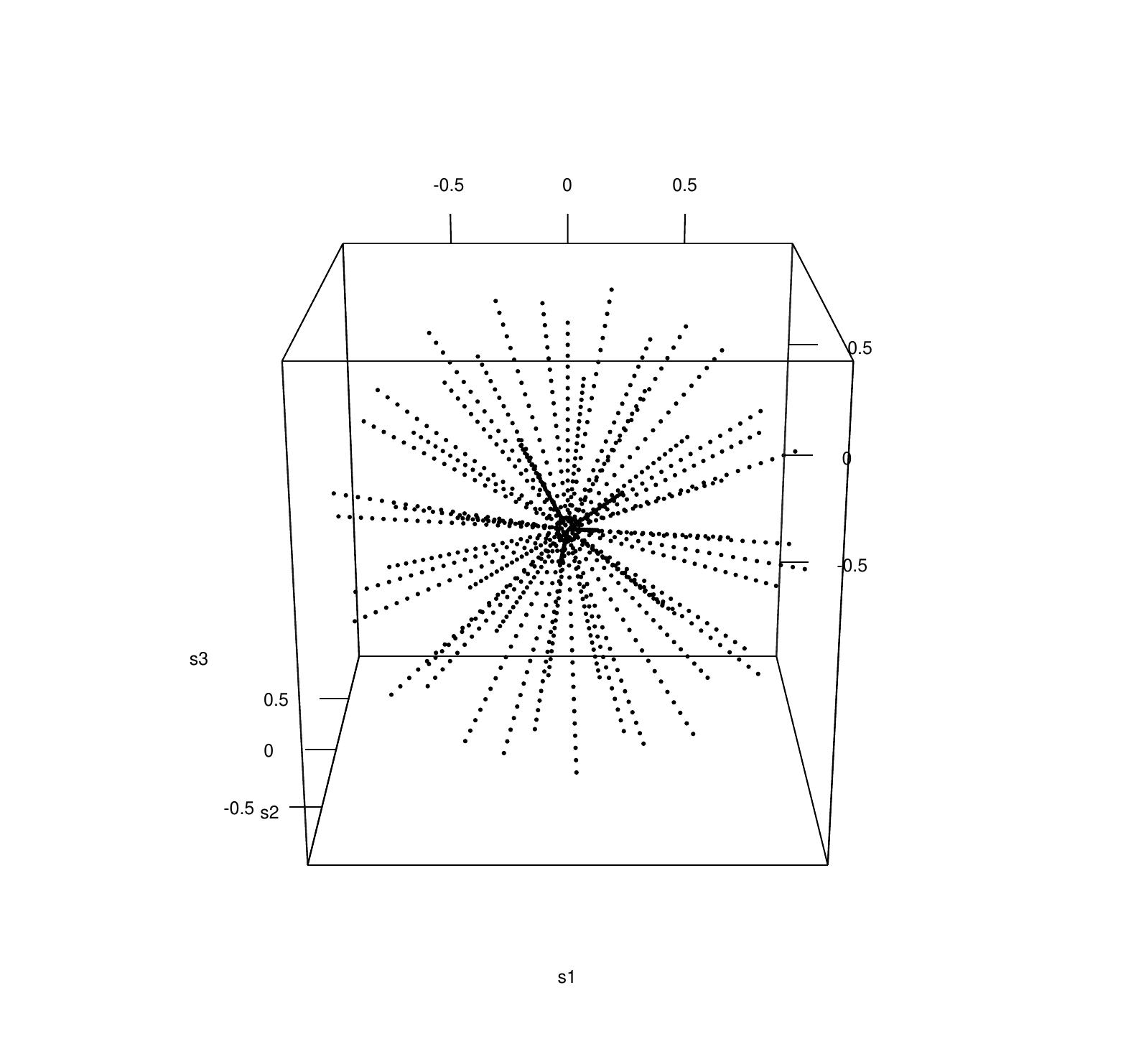}
    \includegraphics[trim=3cm 3cm 3cm 3cm,clip=true,width=6.2cm]{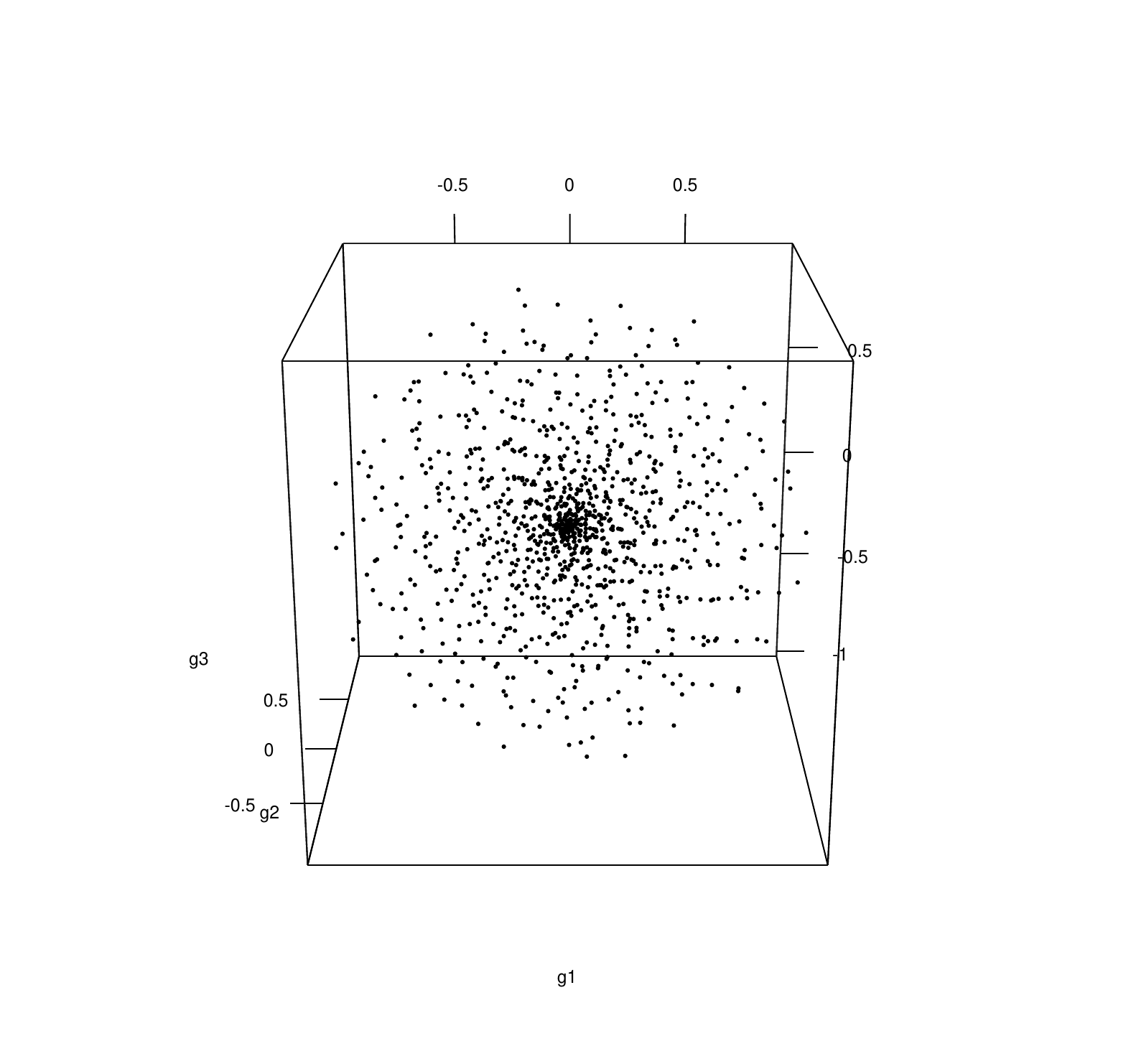}
\caption{A product type grid $\mathcal{P}$ (left panel) and a non-product type grid $\mathcal{N}$ (right panel) for $d=3$.}
\label{ilustrace-3d}
\end{figure}

An example of the two types of grids for $n=144$ in $d=2$ is shown in
Figure~\ref{pokus1-body}. The product type grid (left panel) is constructed as described in the previous paragraph with $n_R=n_S=12$.  The non-product grid (right panel) uses a GLP sequence in $[0,1]^2$ generated by
vector $(1,89)^\top$.
Figure~\ref{ilustrace-3d} provides an example of the two types of grids in $d=3$. Namely, 
the product type grid (left panel) is formed by $n=1100$ points with $n_S=55$ and $n_R=20$, where the unit vectors $\bm{s}_j$ were computed from a GLP sequence in $[0,1]^2$ (generating vector $(1,34)^\top$). 
A non-product grid was obtained as a transformation of $n=1010$ GLP points in $[0,1]^3$ (generating vector $(1,140,237)^\top$).  
Some practical aspects related to  the choice and computation of  the target grids are described in more detail in Section~\ref{sec:4.3}.
We also provide an online example on web page \url{https://www.karlin.mff.cuni.cz/~hudecova/research/mult_perm/Main.html} for both $d=2$ and $d=3$. For $d=3$ the the two types of grids are plotted therein using  3D interactive plots which enable one to rotate them and explore them in more detail.

\subsection{Modifications}\label{sec:3.2}

The CO distribution function $F_{\pm}$ corresponds to the transformation of $P$ to the spherically uniform distribution $\Un_d$.
Instead of $\Un_d$ some authors consider a different reference measure $\mu$, see \cite{galichon2017} or \cite{ghosal}. The corresponding mapping $\widetilde{F}$ then pushes $P$ to $\mu$, and the empirical counterpart of $\widetilde{F}$  naturally requires a different grid $\mathcal{G}_n$. For instance, if $\mu$ is uniform on $[0,1]^d$ then $\mathcal{G}_n$ should be a set of $n$ points ``as uniform as possible'' in $[0,1]^d$.

In Section~\ref{sec:pvalues} we consider a grid  $\mathcal{G}_n^+$ which corresponds to $\mu$ being a restriction of $\Un_d$  on $[0,1]^d$, i.e. a distribution of a random vector $U\cdot \widetilde{\bm{S}}$, where $U$ is uniformly distributed on $[0,1]$ independent of $\widetilde{\bm{S}}$ which is uniformly distributed on $\{\bm x\in[0,1]^d: \|\bm x\|=1\}$.  The   grid $\mathcal{G}_n^+$ can be again chosen in a product form or non-product form,
see Section~\ref{sec:pvalues} for more details.


\section{Application to permutation tests}\label{sec:4}

Let $H_0$ be a null hypothesis which can be tested using a permutation test based on a test statistic $\bm{T}$. Let $\bm{T}_0$ stand for the value of the test statistic computed from the original data, and let 
 $\bm{T}_1,\dots,\bm{T}_{B}$ be the permutation counterparts obtained by a permutation principal analogous to the one described in Section~2.

 In what follows we assume that the distribution of the data as well as the distribution of the test statistic $\bm{T}$ is absolutely continuous, and that the test of interest is two-sided. The latter assumption is  removed in Section~\ref{sec:4.4}.


\subsection{Transportation of the multivariate test statistics}\label{sec:4.1}


Rather than combining the $d$
univariate p-values correspoding to the coordinates of $\bm{T}$, we utilize the center outward 
approach.

\begin{figure}[htb]
\includegraphics[width=10cm]{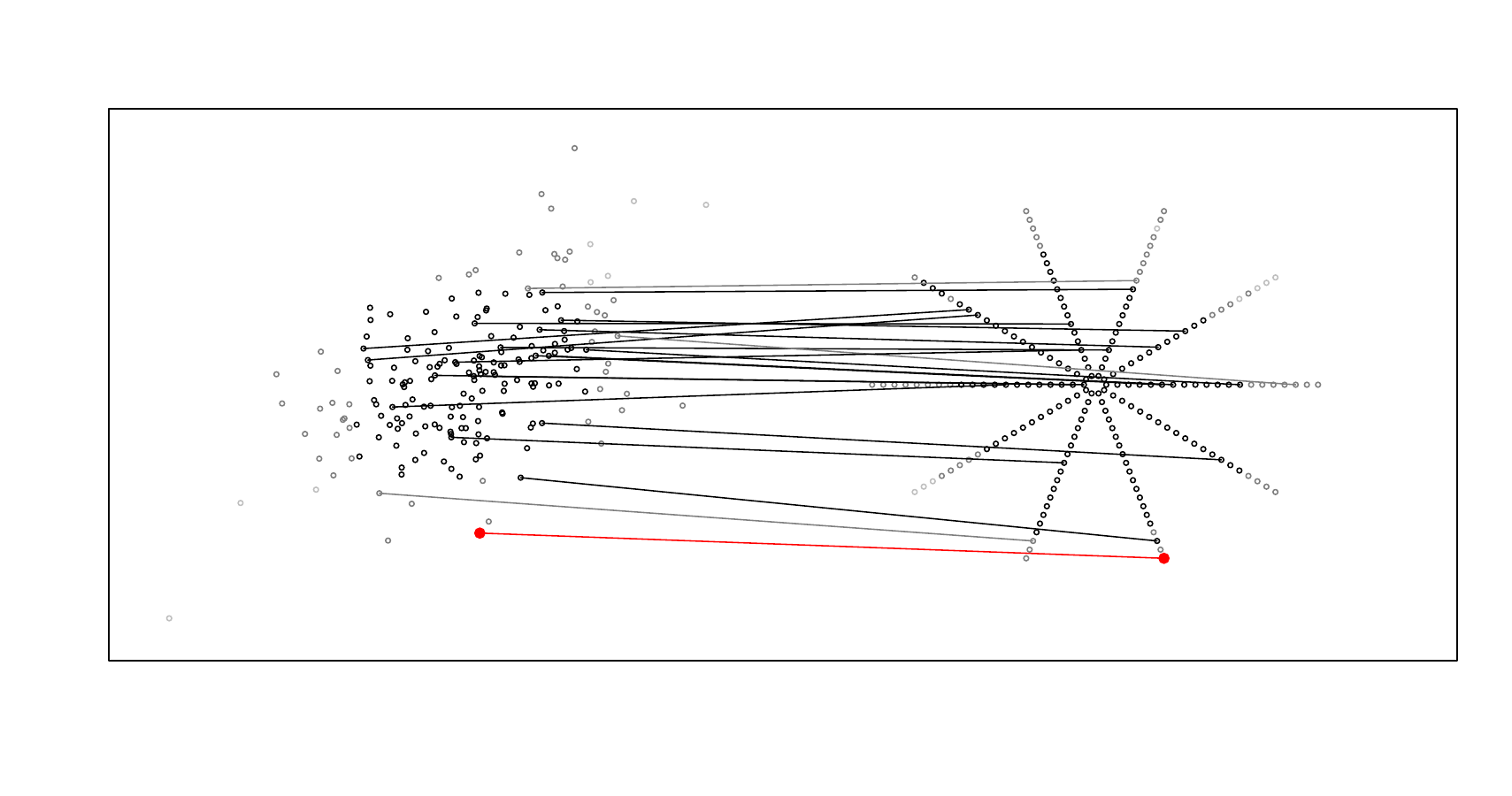}
\caption{The optimal CO transport  $F_{\pm}^{(B)}$ of the observed bivariate test statistic $\bm{T}_0$ (large red point) and the corresponding permutation random variables $\bm{T}_1,\dots,\bm{T}_{B}$, $B=199$.}
\label{fig-papr-trnsp3w}
\end{figure}

Let $\mathcal{G}_B$ be a  grid
of $B+1$ points $\bm{g}_i$, $i=0,\dots,B $, from the unit
ball $\mathbb{S}_d$ and let $\FB_{\pm}$ be the empirical CO distribution function which maps $\bm{T}_0, \bm{T}_1,\dots,\bm{T}_{B}$ to $\mathcal{G}$. 
An example of such mapping $\FB_{\pm}$ in $d=2$ and a
regular product-form grid $\mathcal{G}$ is shown in Figure~\ref{fig-papr-trnsp3w}.

The (standard) bivariate permutation p-value is estimated as the
relative frequency
\begin{equation}\label{eq:pe}
\widehat{p}_e = \frac{1}{B+1} \left(1+\sum_{b=1}^{B}
  \boldsymbol{1}\bigl(\|F_{\pm}^{(B)}(\bm{T}_{b})\| \geq \|F_{\pm}^{(B)}(\bm{T}_{0})\|
  \bigr)\right),
  \end{equation}
  i.e., the relative proportion of points
$\bm{g}_{i}$ that are more distant from the center than the point
$\FB_{\pm}(\bm{T}_0)$. 

Alternatively, the p-value can be approximated also as
\begin{equation}\label{eq:pa}
\widehat{p}_a =1-\|F_{\pm}^{(B)}(\bm T_0)\|,
\end{equation}
i.e. the distance of the transported
(bivariate) test statistic to the unit circle.

The whole procedure is summarized in Algorithm~\ref{alg:alg1}. Some practical aspects of the particular steps 1.--5. are discussed in more detail in Section~\ref{sec:4.5}. 

\begin{algorithm}[bthp]
\caption{Transportation of multivariate test statistics}
\label{alg:alg1}
\begin{algorithmic}[1]
\REQUIRE A $d$-variate two-sided test statistic $\bm{T}_0$ for testing $H_0$.
\STATE Choose the number of permutations $B$. 
\STATE Compute grid $\mathcal{G}_B$ of $B+1$ points.  
\STATE Compute permutation test statistics $\bm{T}_1,\dots,\bm{T}_B$. 
\STATE Compute the transport $\FB_{\pm}$. 
\STATE Calculate the p-value as $\widehat{p}_e$ or $\widehat{p}_a$. 
\ENSURE A scalar p-value of significance of $H_0$ 
\end{algorithmic}
\end{algorithm}

\begin{remark} 

Consider a product type grid $\mathcal{G}_B=\mathcal{P}_B$ with $B+1 = n_R n_S+n_0$. If $n_R$ is large  then the difference between $\widehat{p}_e$ and $\widehat{p}_a$ is negligible, and  
 both $\widehat{p}_e$ and $\widehat{p}_a$ follow under the null hypothesis a distribution that is approximately the uniform distribution over $[0,1]$.

Indeed,  $\widehat{p}_e$ in \eqref{eq:pe} takes values in $\bigl\{\frac{k}{n_R n_S+n_0}, k = n_S, 2n_S,\dots, n_R n_S\bigr\} \cup \{1\}$. Under the null hypothesis the values in the first set are attained with equal probabilities $\frac{n_S}{n_R n_S + n_0}$, and $\widehat{p}_e = 1$ with probability $\frac{n_0}{n_R n_S + n_0}$. Further, $\widehat{p}_a$ in \eqref{eq:pa}  attains values from $\bigl\{\frac{k}{n_R+1}, k=1,\dots,n_R\bigr\}\cup \{1\}$, where all values in the first set are attained with the same probability $\frac{n_S}{n_R n_S + n_0}$ and $\widehat{p}_a = 1$ with probability $\frac{n_0}{n_R n_S + n_0}$.  Since $n_0$ is always chosen negligible compared to $n_R$ and $n_S$, 
 we get the desired distributional properties of $\widehat{p}_e$ and $\widehat{p}_a$. 

In addition, if $n_0 = 0$ then the sets of possible values of $\widehat{p}_e$ and $\widehat{p}_a$ are $\bigl\{\frac{k}{n_R}, k=1,\dots,n_R\bigr\}$ and $\bigl\{\frac{k}{n_R+1}, k=1,\dots,n_R\bigr\}$, respectively, and it holds that 
    \[
     \widehat{p}_a = \widehat{p}_e \frac{n_R}{n_R+1},
    \]
    which shows that their difference is negligible for a large $n_R$.
\end{remark}

\begin{remark}[Handling the ties]
  Although we assume that $\bm T$ has an absolutely continuous distribution,   we cannot completely exclude the occurrence of 
  ties among $\bm{T}_1,\dots,\bm{T}_B$.
  For instance for the three sample problem from Section~\ref{sec:2},
  there are at most
  $\frac{n!}{n_{1}! n_{2}! n_{3}!}$ different results of the permuted
  statistics and we make $B$ permutations, therefore, there is a non-zero
  probability that $\bm{T}_i=\bm{T}_j$ for $i\ne j$. In such a case a randomization approach needs to be used when computing $F_{\pm}^{(B)}$. 
  However, the probability of such event is quite small 
  and our
  practical experiments indicate that the ocurrence of ties can be neglected in practice. 
\end{remark}


\subsection{Transportation of the partial permutation p-values}\label{sec:pvalues}

The approach from the previous section allows us to order the multivariate variables $\bm{T}_0,\bm{T}_1,\dots,\bm{T}_B$ and to compute corresponding empirical quantiles, and the computation of the partial permutation p-values is not required. This section describes an alternative method.

Let $\bm{p}=(p_1,\dots,p_d)^\top$ be the vector of partial permutation p-values computed from $\bm{T}_0$ and let $\bm{p}_b = (p_{1,b},\dots,p_{d,b})$ be the vector of partial p-value corresponding to the components of $\bm{T}_b$, $b=1,\dots,B$, as described in Section \ref{sec:21}. 
Recall that the classical multivariate permutation test uses 
 a combining function $f$ to transform the vector $\bm{p}_b$ to a scalar statistic $Q_b$. Instead of this, we propose to  apply the measure transportation techniques on $\bm{p}_b$. In order to stay consistent with the concept of ``rejecting for large values'', we do not transport the partial p-values directly, but via their complements.
Define
\[
\bm{q}_0=(1-p_1,\dots,1-p_d)^\top, \quad \bm{q}_b=(1-p_{1,b},\dots,1-p_{d,b})^\top.
\]
Then the $j$th partial test statistic is significant if and only if $j$th component of $\bm{q}_0$ is larger than the significance level. Hence, the components of $\bm{q}_0$ behave similarly as test statistics of one sided tests, so it is not suitable to transport the points $\bm{q}_b$, $b=0,\dots,B$ to the same grid as in Section~\ref{sec:4.1}. Instead, we will make use of a regular grid ${\mathcal{G}}_B^+$ of $B+1$ points in 
the set $\{\bm{x}\in[0,1]^d: \|\bm{x}\|\leq 1\}$. Namely, we use two types of grids analogous to those from Section~\ref{sec:grids}:
\begin{itemize}
\item a product form grid, abbreviated as $\mathcal{P}^+$, which consists of points $\bm{g}_{ij}=\frac{i}{n_R+1} \bm{s}_j$ for $i=1,\dots,n_R$, and $j=1,\dots,n_S$,  $B+1=n_R n_S$, where  $\bm{s}_j$ are computed as a suitable transformation of a low-discrepancy sequence in $[0,1]^{d-1}$ to the set $\{\bm{x}\in[0,1]^d: \|\bm{x}\|= 1\}$,
\item a non-product form grid, denoted as $\mathcal{N}^+$, consisting of directly transformed $B+1$ low-discrepancy points in $[0,1]^d$ to the set $\{\bm{x}\in[0,1]^d: \|\bm{x}\|\leq 1\}$. 
\end{itemize}
Analogously as in Section~\ref{sec:3}, one calculates $\widetilde{F}_{\pm}^{(B)}$ as a mapping of $\bm{q}_0,\bm{q}_1,\dots,\bm{q}_B$ to ${\mathcal G}_B^+$, which minimizes the squared Euclidean distances.  
The resulting p-value of the
permutation test of $H_0$ can be then computed as  
\[
\widetilde{p}_e = \frac{1}{1+B}\sum_{b=1}^{B}
\boldsymbol{1}\left(\|\widetilde{F}_{\pm}^{(B)}(\bm{q}_b)\| \geq \|\widetilde{F}_{\pm}^{(B)}(\bm{q}_0) \| \right),
\]
or simply as $\widetilde{p}_a=1-\|\widetilde{F}_{\pm}^{(B)}(\bm{q}_0)\|$.

\begin{figure}
\includegraphics[width=10cm]{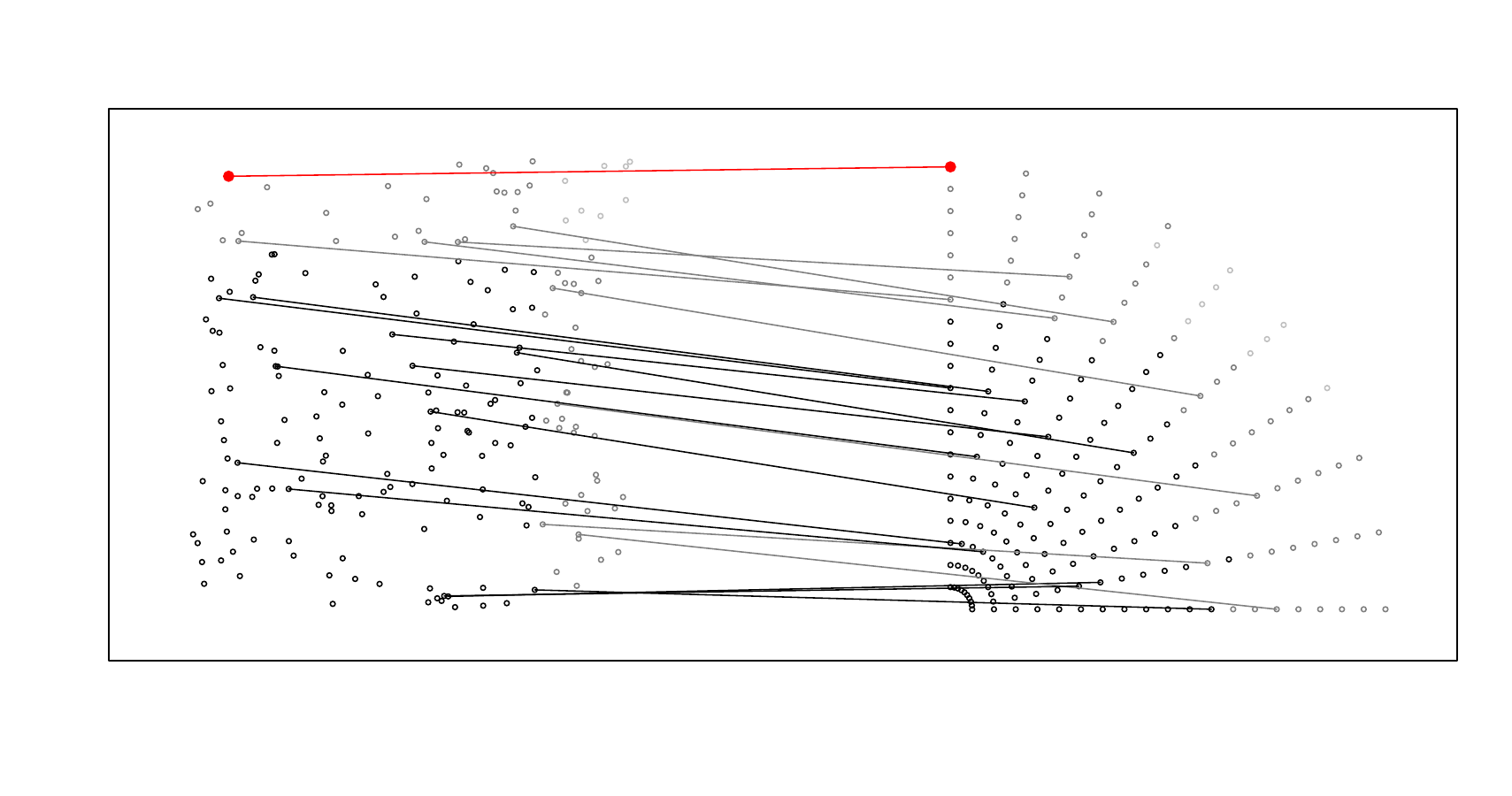}
\caption{Transport of complements $\bm{q}_b$, $b=1,\dots,B$, of vectors of partial permutation p-values onto a product grid $\mathcal{P}^+_B$ of $B+1=200$ points.}\label{fig-papr-trnsp3pval}
\end{figure}

An example of a transport $\widetilde{F}_{\pm}^{(B)}$ of $\bm{q}_b$, $b=0,\dots,B$, 
obtained in the 
simulated example in Section~\ref{sec:21} to a product form grid
is plotted in   Figure~\ref{fig-papr-trnsp3pval} for $B+1=200$. The whole procedure is summarized in Algorithm~\ref{alg:alg2}, while some practical aspects on the particular steps are postponed to Section~\ref{sec:4.5}.

\begin{algorithm}[bthp]
\caption{Transportation of multivariate permutation p-values}
\label{alg:alg2}
\begin{algorithmic}[1]
\REQUIRE A $d$-variate test statistic $\bm{T}_0$ for testing $H_0$.
\STATE Choose the number of permutations $B$. 
\STATE Compute grid ${\mathcal{G}}_B^+$ of $B+1$ points.  
\STATE Compute permutation test statistics $\bm{T}_1,\dots,\bm{T}_B$.
\STATE Compute the partial permutation p-values and the corresponding vectors of complements $\bm{q}_j$, $j=0,1,\dots,B$. 
\STATE Compute the transport $\widetilde{F}_{\pm}^{(B)}$. 
\STATE Calculate the p-value as $\widetilde{p}_e$ or $\widetilde{p}_a$. 
\ENSURE A scalar p-value of significance of $H_0$ 
\end{algorithmic}
\end{algorithm}

\subsection{Interpretation of partial contributions}\label{sec:4.3}

Unlike the multivariate permutation tests, the optimal transport approach provides also additional information on the importance of the marginal hypotheses. Consider first the approach from Section~\ref{sec:4.1}, where the test statistics $\bm{T}_0,\dots,\bm{T}_B$ are transported via mapping $F_{\pm}^{(B)}$.

Apart from the p-value computed from $F_{\pm}^{(B)}(\bm T_0)$, the direction vector
\[
\bm{D}=(D_1,\dots,D_d)^\top = \frac{F_{\pm}^{(B)}(\bm T_0)}{\| F_{\pm}^{(B)}(\bm T_0)\|}
\]
can  reveal how the partial statistics $T_{j,0}$, $j=1,\dots,d$, contribute to the rejection of $H_0$. 
As mentioned in Section~\ref{sec:2},  $(1-\widehat{p}_a)^2$ can be interpreted as a non-conformity score which measures the devation from the null hypothesis. It follows  from \eqref{eq:pa} that $\| F_{\pm}^{(B)}(\bm T_0)\|^2=(1-\widehat{p}_a)^2$, so we obtain a decomposition
\[
(1-\widehat{p}_a)^2=\| F_{\pm}^{(B)}(\bm T_0)\|^2=\sum_{j=1}^d F_{\pm,j}^{(B)}(\bm{T}_0)^2,
\]
where $F_{\pm}^{(B)}(\bm T_0)=\left(F_{\pm,1}^{(B)}(\bm{T}_0),\dots,F_{\pm,d}^{(B)}(\bm{T}_0) \right)^\top$.
Hence, $F_{\pm,j}^{(B)}(\bm{T}_0)^2$ can be interpreted as an absolute contribution of the $j$-th partial test statistic to the rejection of the composite null hypothesis  because the null hypothesis is not rejected if the sum of these contributions is too small. When the null hypothesis is rejected, it makes sense to consider
\begin{equation}\label{eq:contr}
(D_1^2,\dots,D_d^2) = \frac{\Bigl(F_{\pm,1}^{(B)}(\bm{T}_0)^2,\dots,F_{\pm,d}^{(B)}(\bm{T}_0)^2\Bigr)}{\|F_\pm^{(B)}(\bm{T}_0)\|^2}
\end{equation}
that can be  interpreted
as  a relative contribution of the $j$-th partial test statistic in terms of percentages because  $\sum_i D_i^2 = 1$.

Similarly, if the method from Section~\ref{sec:pvalues} is used and the partial permutation p-values are transported via $\widetilde{F}_{\pm}^{(B)}$, then one can define the vector of individual
contributions as
\[
 \frac{\Bigl(\widetilde{F}_{\pm,1}^{(B)}(\bm{q}_0)^2,\dots,\widetilde{F}_{\pm,d}^{(B)}(\bm{q}_0)^2\Bigr)}{\|\widetilde{F}_\pm^{(B)}(\bm{q}_0)\|^2}.
\]

%

\smallskip

We illustrate this concept on the three sample example from Section~\ref{sec:2}, where $H_{0,1}:\mu_1=\mu_2$ and $H_{0,2}:\mu_1=\mu_3$, and $T_1$ and $T_2$ are the standard $t$-statistics. 

\begin{figure}[htb]
\includegraphics[width=10cm]{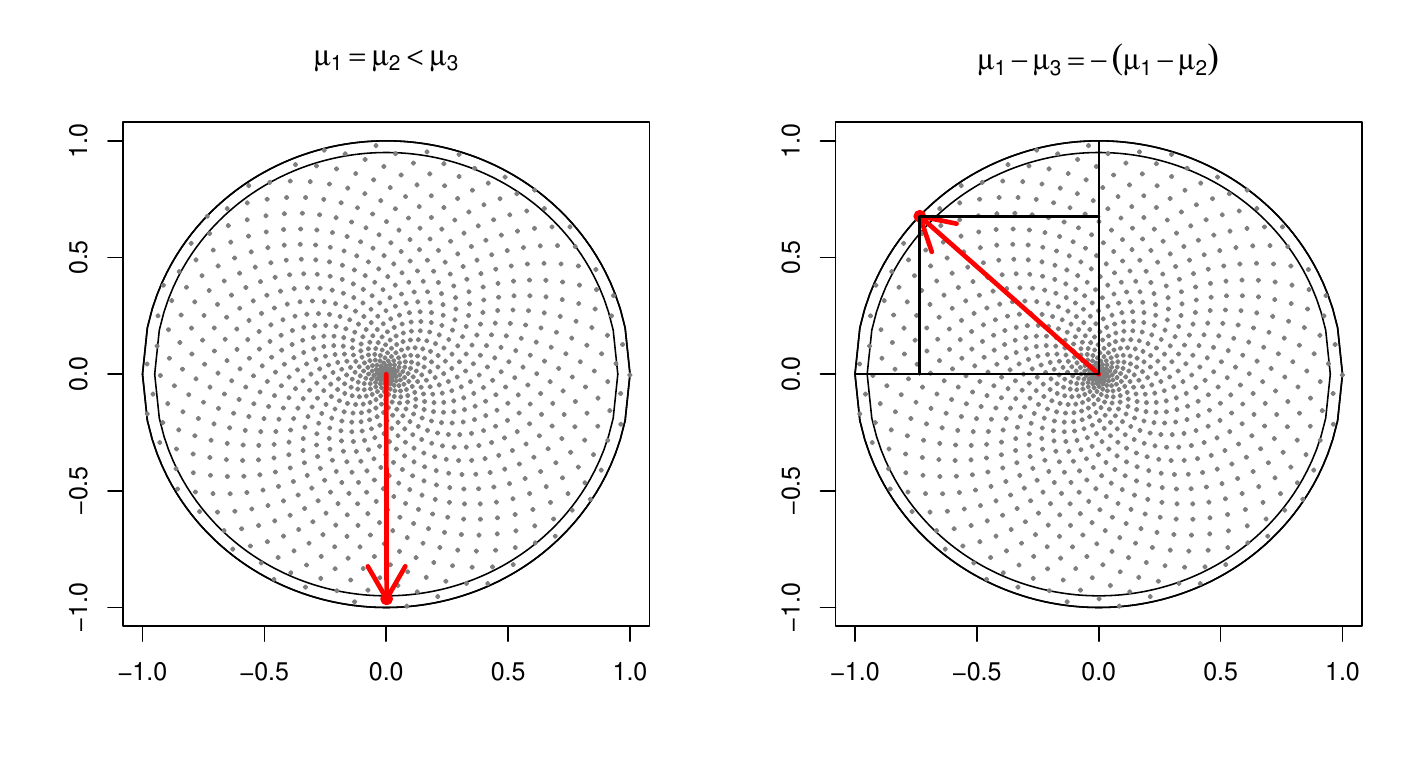}\\
\includegraphics[width=10cm]{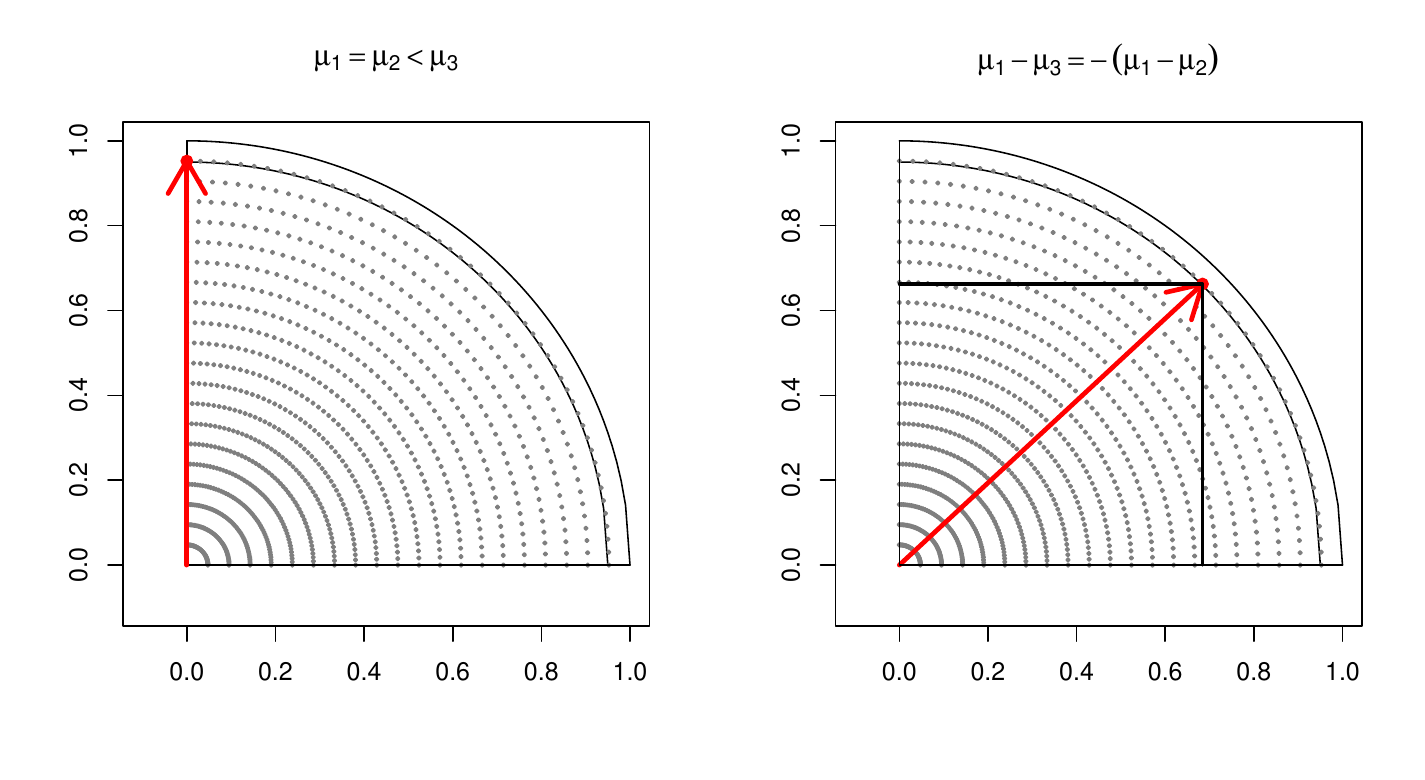}
\caption{Decomposition of a significant test statistic in two setups, with $F_\pm^{(B)}(\mathbf{T}_0)$ denoted by a red arrow, into directions and partial contributions: as an illustration, two-sided test statistics and a non-product set $\mathcal{N}$ are used in the first two plots (first row), partial permutation p-values and a product grid $\mathcal{P}^+$ are plotted in the second row.}
\label{ilustrace-prpi1}
\end{figure}

Suppose first that $H_{0,1}$ holds and $\mu_1<\mu_3$. Then we expect $T_{1,0}$ to be close to zero, and $T_{2,0}$ to be large negative. On the other hand, if $T_{j,b}$, $j=1,2$ are computed from a permuted sample, then it is likely that $|T_{1,b}|>|T_{1,0}|$  and  $T_{2,0}<T_{2,b}$. 
 Hence, the point $\bm{T}_0=(T_{1,0},T_{2,0})^\top$ is expected to be transported to a grid point close to $(0,-1)$. The corresponding vector of individual contributions would be $(0,1)$ indicating that the two hypotheses contribute by 0 \% and 100 \% respectively. 
This situations is illustrated in
 an idealized example in the left upper panel of Figure~\ref{ilustrace-prpi1}
Here,  the red point corresponds to the transported
observed test statistic $F_\pm^{(B)}(\mathbf{T}_0)$  and it lies far from the
centre in the direction of the violated marginal null hypothesis. As described
previously, we reject the null hypothesis if  
$\|F_\pm^{(B)}(\mathbf{T}_0)\|>1-\alpha$. At the same time, the ``downward'' 
pointing arrow indicates that the rejection is due mainly to the second component 
of the composite null hypothesis. 

Similar considerations hold also for the transportation of partial permutation p-values, which are, for this example, expected to be $p_1\approx 1$ and $p_2$ small, leading to a complement vector $\bm{q}_0$ close to $(0,1)$. Such point is then transported to $\widetilde{F}_{\pm}^{(B)}(\bm{q}_0)$ close to $(0,1)$, see the lower left panel of Figure~\ref{ilustrace-prpi1}. The red arrow again indicates how the individual tests contribute to the rejection of the null hypothesis.

\smallskip

Suppose now a different situation where $\mu_1-\mu_3=\gamma=-(\mu_1-\mu_2)$ for some $\gamma>0$, so the two marginal hypotheses  contribute equally to the overall decision. The test statistic $T_{1,0}$ is expected to be large negative and $T_{2,0}$ to be large positive. 
An idealized example of a transport of such point $\bm{T}_0$ is provided  in the upper right plot in Figure~\ref{ilustrace-prpi1}.
Here, the direction $\bm{D}$ of $\FB_{\pm}(\bm{T}_0)$ is $(-1/\sqrt{2},1/\sqrt{2})$, so the
 vector of individual contributions is $(1/2,1/2)$. The vector of p-values for this example is expected to be close to $(0,0)$, so the complement vector $\bm{q}_0$ is close to the extreme $(1,1)$. This point is then transported to a grid point close to $(1/\sqrt{2},1/\sqrt{2})$. 

 \smallskip
 
In practice, the point $\bm{T}_0$ is not transported exactly as shown in the upper two panels of Figure~\ref{ilustrace-prpi1}
but $\FB_{\pm}(\bm{T}_0)$ is expected to be somewhere near  the idealized point, depending on $B$ and the chosen grid $\mathcal{G}_B$. The same holds for the transport of $\bm{p}_0$.
However, a simulation study presented in Section~\ref{sec:decomp} reveals that the vector of individual contributions is on average very close to the true vector of contributions for both methods.



\subsection{Modification for ``mixed'' alternatives}\label{sec:4.4}


Recall that Section~\ref{sec:4.1} assumed that all the partial problems are two-sided.  
Let $\bm T$ be again a $d$-dimensional test statistic for testing $H_0$. If all the partial tests are one-sided such that the corresponding hypothesis is rejected for large values, then the approach from Section~\ref{sec:4.1} can be easily adapted to this situation 
by considering a grid $\mathcal{G}_B^+$ from 
 Section~\ref{sec:pvalues}.


More generally, assume that   the partial tests  reject for $|T_j|$ large (two-sided tests) for $j\in\mathcal{J}\subset\{1,\dots,d\}$, while other partial tests reject for $T_j$ large (one-sided tests) for $j\in\{1,\dots,d\} \setminus \mathcal{J}$. 
The multivariate permutation test from Section~\ref{sec:4.1} can be  then adapted to this situation simply by considering  
 a regular set of $B+1$ points $\widetilde{\mathcal{G}}_B$ in $\left\{\bm{x}: \|\bm{x}\|\leq 1, \ x_j\geq 0 \text{ for } j\in\{1,\dots,d\} \setminus \mathcal{J}\right\}$. Construction of such  grid   is essentially an analogue to
the construction of $\mathcal{G}_B$ and ${\mathcal{G}}_B^+$.

\subsection{Practical implementation}\label{sec:4.5}

Let us discuss  the steps  1., 2. and 4. of Algorithm~\ref{alg:alg1} in more detail.
A particular examples in R software \citep{R} are provided online, on web page \url{https://www.karlin.mff.cuni.cz/~hudecova/research/mult_perm/Main.html}.

{\it Step 1.} It is common to take $B\approx 1\,000$ or larger in permutation tests. In our applications the choice of $B$ is also related to the choice of the grid type, see Section~\ref{sec:grids}. 
If one wishes to use a  product form grid then it is suitable to   
chose $B$ such that $B+1=n_R n_S$ or $B+1=n_R n_S+1$ (so $n_0$ is 0 and 1 respectively).  
The decomposition $n_R$ and $n_S$ should be chosen suitably with respect to the dimension $d$.
Moreover, the smallest possible p-value $\widehat{p}_a$ is $1/(n_R+1)$, so one has to take $n_R\geq 1/\alpha-1$, where $\alpha$ is the chosen significance level. For $\alpha=0.05$, this gives $n_R\geq 19$. Note that if $n_R\in[19,38]$ then the null hypothesis is rejected only if $\FB_{\pm}(\bm{T}_0)$ lies on the most outer circle of the grid, i.e. $\|\FB_{\pm}(\bm{T}_0)\|=n_R/(n_R+1)$. 

If a non-product grid is going to be used there is no need for factorization into a product $n_Rn_S$, but it is customary to take $B$ such that $(B+1)\cdot \alpha$ is an integer, where $\alpha$ is the chosen significance level. In that case we reject for if the value of $\bm{T_0}$ is transported to one of the $(B+1)\cdot \alpha$ most extreme points of the grid (see also the online example). 
Moreover, if one wishes to use a GLP set to generate the $B+1$ low-discrepancy points in $[0,1]^d$ then 
 $B$ should be chosen with respect to this fact, because there are some practical limitations, see below.   
 

\smallskip

{\it Step 2.} 
For $d>2$ both types of grids require a construction of a low-discrepancy sequence in $[0,1]^r$ for either $r=d-1$ or $r=d$. As mentioned in Section~\ref{sec:grids}, this sequence can be 
obtained by either a Halton method or a Sobor method,
which are both available in \textsf{R}  in
package \texttt{randtoolbox} \citep{randtool}. 
As an alternative construction, used also in some examples in this paper, one can compute
the so-called good lattice point (GLP) set described in \citet[Section 1.3]{fang}, where the lattice points set of size $B+1$ in $[0,1]^d$ is obtained by calculating a set of properly rescaled modulo $(B+1)$ multiples of a selected generating vector $h=(h_1,\dots,h_d)^\top$. 
It can be shown that, for certain values of $B$, appropriately chosen generating vector leads to a well uniformly scattered set of points \citep{korobov1959computation,hlawka1962angenaherten}.
A table of optimal generating vectors for various $B$ and $d$ is given, e.g., in 
\citet[Appendix~A]{fang}. This means that a GLP sequence of length $B+1$ with low discrepancy in $[0,1]^d$ can be easily generated only for some particular values of $B+1$. However, this should not be seen as a serious restriction because the number of permutations $B$ can be chosen quite flexibly.

 

\smallskip

{\it Step 4.} For a given grid set $\mathcal{G}_B$  the  transportation  $F_{\pm}^{(B)}$ can be  computed using so called Hungarian algorithm \citep{papad} which is implemented for instance in package \texttt{clue} in \textsf{R}.

\section{Simulation study} \label{sec:sim}

Let us now investigate the concept in a small simulation study. We start with a simple example, similar to the simulated example from Section~\ref{sec:21}.

\subsection{Testing for means under homoscedasticity.} Consider a homoskedastic three sample problem, where we generate independent observations $X_{j,i}\sim N(\mu_j,\sigma^2)$, $i=1,\dots,n_j$ for samples $j\in\{1,2,3\}$ and for $n_1=n_2=n_3$.

Our three basic simulation setups are as follows:
\begin{description}
\item[$H_0$] $\mu_1=\mu_2=\mu_3=0$,
\item[$A_\delta$] $\mu_1=\mu_2=0$, $\mu_3=\delta$, 
\item[$B_\delta$] $\mu_1=0$, $\mu_2=\delta/2=-\mu_3$, 
\end{description}
with the parameter $\delta\in\{1,2\}$, $\sigma^2=1$, $n_1 \in \{5,10\}$. 

In this setup, it can be expected that very good results will be achieved by the
usual F-test. Following the example from Section~\ref{sec:2}, the F-test will be compared to the multivariate
permutation test  (Algorithm 1)  based on the optimal transportation of the bivariate t-test statistic $\bm T=(T_1,T_2)^\top$, where $T_1$ compares the first
and the second sample and $T_2$ compares the first and the third sample.
 Next to this, we also compute the optimal transport and the corresponding p-value of the one-sided test statistic $\widetilde{\bm T}=(|T_1|,|T_2|)^\top$, 
and the transport of the vector of partial permutation p-values (Algorithm 2). 
The result are compared to the classical multivariate
permutation tests utilizing the most commonly used Tippett, Fisher, and Liptak combining functions
described in Section~\ref{sec:2}. More precisely, we compare the following tests:
\begin{description}
\item[$F$] usual F-test (based on $F$ quantiles),
\item[$c(T)$] multivariate permutation test with the Tippett combining function,
\item[$c(F)$] multivariate permutation test with the Fisher combining function,
\item[$c(L)$] multivariate permutation test with the Liptak combining function,
\item[$t(\mathcal{P})$]  transport of the two-sided test statistic $\bm T$ to a product grid $\mathcal{P}$,
\item[$t(\mathcal{N})$]  transport of the two-sided test statistics $\bm T$ to a non-product grid $\mathcal{N}$,
\item[$t(\mathcal{P}^+)$]  transport of the one-sided test statistics $\widetilde{\bm T}$ to the positive  product grid $\mathcal{P}^+$,
\item[$t(\mathcal{N}^+)$]  transport of the one-sided test statistics $\widetilde{\bm T}$ to the positive non-product grid~$\mathcal{N}^+$ ,
\item[$p(\mathcal{P}^+)$]  transport of the vector of partial permutation p-values to $\mathcal{P}^+$,
\item[$p(\mathcal{N}^+)$]  transport of the vector of partial permutation p-values to $\mathcal{N}^+$.
\end{description}

\begin{table}
\begin{tabular*}{\textwidth}{lr@{\extracolsep{\fill}}r@{\extracolsep{\fill}}r@{\extracolsep{\fill}}r@{\extracolsep{\fill}}r@{\extracolsep{\fill}}r@{\extracolsep{\fill}}r@{\extracolsep{\fill}}r@{\extracolsep{\fill}}r@{\extracolsep{\fill}}r@{\extracolsep{\fill}}r} 
\hline\hline
  & $n_1$ & $F$&$c(T)$&$c(F)$&$c(L)$&$t(\mathcal{P})$&$t(\mathcal{N})$&$t(\mathcal{P}^+)$&$t(\mathcal{N}^+)$&$p(\mathcal{P}^+)$&$p(\mathcal{N}^+)$ \\ 
 \cline{2-2}\cline{3-6}\cline{7-12}
 &5&4.1&4.4&5.3&5.7&3.5&4.8&5.0&5.6&4.9&5.4 \\ 
 \raisebox{1.5ex}[0pt]{$H_0$}&10&6.4&4.9&5.1&4.1&4.8&4.4&5.2&5.7&6.1&4.7 \\ 
\cline{2-2}\cline{3-3}\cline{4-6}\cline{7-8}\cline{9-10}\cline{11-12}
&5&{25.5}&22.6&16.0&14.4&\textbf{24.4}&\textbf{24.2}&18.6&19.7&20.4&19.2 \\ 
 \raisebox{1.5ex}[0pt]{$A_1$}&10&{57.1}&\textbf{48.8}&35.6&19.5& \textbf{52.5}&47.9&45.9&46.2&43.4&45.3 \\ 
\cline{2-2}\cline{3-3}\cline{4-6}\cline{7-8}\cline{9-10}\cline{11-12}
 &5&{84.2}&67.0&50.8&33.8&\textbf{72.3}&\textbf{72.5}&63.7&65.7&68.3&63.5 \\ 
 \raisebox{1.5ex}[0pt]{$A_2$}&10&{99.4}&97.4&94.4&60.6&\textbf{99.1}&\textbf{98.1}&97.0&97.4&96.2&94.5 \\ 
 \cline{2-2}\cline{3-3}\cline{4-6}\cline{7-8}\cline{9-10}\cline{11-12}
 &5&{24.5}&13.2&7.5&7.0&\textbf{20.6}&\textbf{22.5}&10.5&11.1&13.1&11.3 \\ 
 \raisebox{1.5ex}[0pt]{$B_1$}&10&{45.1}&24.8&19.3&11.2&\textbf{43.0}&\textbf{44.0}&20.2&20.7&22.6&23.1 \\ 
 \cline{2-2}\cline{3-3}\cline{4-6}\cline{7-8}\cline{9-10}\cline{11-12}
 &5&{70.2}&36.3&31.8&31.5&\textbf{65.1}&\textbf{64.1}&34.6&34.5&34.5&34.2 \\ 
 \raisebox{1.5ex}[0pt]{$B_2$}&10&{97.1}&76.0&79.6&76.1&\textbf{95.4}&\textbf{95.2}&76.3&76.6&70.5&70.0 \\ 
 \hline\hline 
 \end{tabular*}
\caption{Empirical size and power (in \%) of multivariate permutation tests using combining functions and optimal transport in a three sample testing problem, sample sizes $n_1=n_2=n_3$, $n_1\in \{5,10\}$, homoskedastic case with $\sigma^2=1$, significance level $\alpha=0.05$, 1000 simulations, bold font denotes two most powerful permutation tests in each simulation setup.}\label{tabka1}
\end{table}

For all variants of the multivariate permutation test, we use $B=999$ permutations with the exception of 
the non-product grid,  because  its construction is based on a
GLP set of $B+1$ points, and it is described in
Section~\ref{sec:4.5} that only some specific choices of $B+1$ are possible in this case. 
Hence, for a non-product grid we  set $B+1=987$ and the GLP sequence in $[0,1]^2$ is 
 computed using generating vector $h=(1,610)^\top$, as
recommended in \citet[Appendix~A]{fang}. The product form grid was
generated with $n_0=0$, $n_R=20$, and $n_S=50$ for the two-sided test statistic and
$n_R=40$ and $n_S=25$ for the one-sided tests. In the two-sided setup, the null
hypothesis is rejected on level $\alpha=0.05$ if and only if the transported test
statistics falls in the outermost ring (with p-value equal to $1/21 \doteq 0.0476$).
In the one-sided setup, the null hypothesis is rejected on level $\alpha=0.05$ if the test statistic
falls in the two outermost rings, i.e., for p-value $1/41$ or $2/41$.

Simulation results are summarized in Table~\ref{tabka1}. For all tests, the empirical level is reasonably close to the nominal level $\alpha=0.05$. Looking at the simulation setup $A_\delta$, the highest power is always achieved by the F-test, followed closely by the optimal transport permutation tests using the two-sided test statistics. Somewhat smaller power is obtained by looking at the univariate test statistics or the p-values (using either the Fisher combining function or the optimal transport approach). 
Interestingly, the advantages of using the two-sided test statistics are even clearer in the simulation setup $B_\delta$, where the largest difference (between groups 2 and 3) is not considered directly in the test statistics $T_1$ and $T_2$. 

\subsection{Decomposition to partial contributions} \label{sec:decomp}

As described in Section~\ref{sec:4.3}, 
 the optimal transport approach provides additional information on the importance of the marginal hypotheses, which is a crucial difference compared to  the F-test or the multivariate permutation tests. For $d=2$ the vector of individual contributions is $(D_1^2,D_2^2)$, so it suffices to look at one component of this vector. Alternatively, one can look at the angle between the $x$ axis and the vector $(D_1,D_2)$, i.e. $\varphi$ such that $D_1=\cos \varphi$ and $D_2=\sin \varphi$.

In order to elucidate this concept, the simulated average angles $\varphi$ and partial contributions $D_2^2\cdot 100 \%$ (percentage contribution of $T_2$) are reported in Table~\ref{tabka1-prpi10}.
Recall that $T_1$ tests $H_{0,1}:\mu_1=\mu_2$, while $T_2$ corresponds to $H_{0,2}: \mu_1=\mu_3$. In $A_{\delta}$ setup $H_{0,1}$ holds, so only $T_2$ should contribute to the rejection, and the idealized value of $(D_1,D_2)$  is $(0,1)$ corresponding to $\varphi=3\pi/2$ and $D_2^2=1$, see also the top left panel of Figure~\ref{ilustrace-prpi1}.  This is confirmed in Table~\ref{tabka1-prpi10} by values in columns corresponding to $t(W)$ and $t(G)$. If the absolute values $|T_j|$, $j=1,2$, are considered,  then 
$(D_1,D_2)$ is expected to be close to $(0,1)$,
corresponding to the angle $\pi/2$. The empirical counterparts in columns $t(\mathcal{P}^+)$, $t(\mathcal{N}^+)$ seem to be slightly further from the expected quantities, and the same conclusion is obtained for the transport of the multivariate p-values, reported in columns $p(\mathcal{P}^+)$ and $p(\mathcal{N}^+)$, where the agreement with the expected quantities seems to be the worst. Also, the larger $\delta$, the closer the empirical quantities are to the nominal ones. 

The setting $B_{\delta}$ corresponds to the idealized situation plotted in the two right panels of Figure~\ref{ilustrace-prpi1}. So for the transport of $(T_1,T_2)^\top$, $(|T_1|,|T_2|)^\top$ and $(p_1,p_2)^\top$ the expected angles are $3\pi/4$, $\pi/2$ and $\pi/2$ respectively, and the contribution of $T_2$ is expected to be 50 \%. The sample analogous in Table~\ref{tabka1-prpi10} are very close to these nominal values for all three cases. Here the difference between results for $\delta=1$ and $\delta=2$ is negligible.


To sum up, 
the results under both alternatives confirm that the direction $\bm{D}$ of the transported
observed test statistic $F_\pm^{(n)}(\mathbf{T}_0)$ really may be used as a tool for uncovering the
underlying reasons for rejecting the composite null hypothesis.

\begin{table}
\begin{tabular*}{\textwidth}{*{1}{l}@{\extracolsep{\fill}}r@{\extracolsep{\fill}}r@{\extracolsep{\fill}}r@{\extracolsep{\fill}}r@{\extracolsep{\fill}}r@{\extracolsep{\fill}}r@{\extracolsep{\fill}}r@{\extracolsep{\fill}}r@{\extracolsep{\fill}}r@{\extracolsep{\fill}}r@{\extracolsep{\fill}}r@{\extracolsep{\fill}}r} 
\hline\hline
& \multicolumn{2}{c}{$t(\mathcal{P})$}&\multicolumn{2}{c}{$t(\mathcal{N})$}&\multicolumn{2}{c}{$t(\mathcal{P}^+)$}&\multicolumn{2}{c}{$t(\mathcal{N}^+)$}&\multicolumn{2}{c}{$p(\mathcal{P}^+)$}&\multicolumn{2}{c}{$p(\mathcal{N}^+)$} \\ 
 \cline{2-13}
 $A_1$&95\%&1.57$\pi$&94\%&1.58$\pi$&91\%&0.40$\pi$&91\%&0.40$\pi$&88\%&0.39$\pi$&88\%&0.39$\pi$ \\ 
 $A_2$&98\%&1.55$\pi$&98\%&1.54$\pi$&98\%&0.46$\pi$&98\%&0.45$\pi$&93\%&0.41$\pi$&93\%&0.41$\pi$ \\ 
 \cline{2-5}\cline{6-9}\cline{10-13}
 $B_1$&49\%&0.75$\pi$&49\%&0.75$\pi$&50\%&0.25$\pi$&51\%&0.25$\pi$&51\%&0.25$\pi$&50\%&0.25$\pi$ \\ 
 $B_2$&50\%&0.75$\pi$&49\%&0.75$\pi$&49\%&0.25$\pi$&51\%&0.25$\pi$&48\%&0.24$\pi$&50\%&0.25$\pi$ \\ 
\hline\hline
 \end{tabular*}\caption{Percentage contribution of the second  hypothesis and angles $\varphi$  of $F_\pm^{(n)}(\mathbf{T}_0)$. The averages are computed over significant simulation results, significance level $\alpha=0.05$, $n_1=10$ observations in each sample.}\label{tabka1-prpi10}
\end{table}


\subsection{Testing for means under heteroscedasticity} \label{sec:sim:het}

\begin{table}
\begin{tabular*}{\textwidth}{lr@{\extracolsep{\fill}}r@{\extracolsep{\fill}}r@{\extracolsep{\fill}}r@{\extracolsep{\fill}}r@{\extracolsep{\fill}}r@{\extracolsep{\fill}}r@{\extracolsep{\fill}}r@{\extracolsep{\fill}}r@{\extracolsep{\fill}}r@{\extracolsep{\fill}}r} 
 \hline \hline
  & $n_1$ & $F$&$c(T)$&$c(F)$&$c(L)$&$t(\mathcal{P})$&$t(\mathcal{N})$&$t(\mathcal{P}^+)$&$t(\mathcal{N}^+)$&$p(\mathcal{P}^+)$&$p(\mathcal{N}^+)$ \\ 
 \cline{2-2}\cline{3-12}
&5&10.1&7.5&7.7&5.6&\textbf{11.8}&\textbf{11.7}&10.9&9.0&9.9&8.5 \\ 
 \raisebox{1.5ex}[0pt]{$H_0$}&10&{9.5}&7.6&6.6&5.2&\textbf{8.0}&\textbf{9.2}&7.8&6.6&7.8&7.6 \\ 
\cline{2-2}\cline{3-3}\cline{4-6}\cline{7-8}\cline{9-10}\cline{11-12}
 &5&14.2&14.3&10.5&6.0&\textbf{14.4}&\textbf{17.0}&12.4&10.8&13.0&13.7 \\ 
 \raisebox{1.5ex}[0pt]{$A_1$}&10&{18.0}&11.7&9.3&8.1&\textbf{16.8}&\textbf{16.9}&13.4&14.1&13.7&12.9 \\ 
\cline{2-2}\cline{3-3}\cline{4-6}\cline{7-8}\cline{9-10}\cline{11-12}
 &5&{23.5}&\textbf{22.7}&18.6&12.2&21.7&\textbf{23.8}&20.6&20.9&21.1&22.1 \\ 
 \raisebox{1.5ex}[0pt]{$A_2$}&10&{39.8}&29.5&25.8&16.4&\textbf{33.2}&\textbf{33.2}&29.0&28.3&26.6&26.4 \\ 
\cline{2-2}\cline{3-3}\cline{4-6}\cline{7-8}\cline{9-10}\cline{11-12}
 &5&12.0&13.7&14.7&7.4&\textbf{20.6}&\textbf{20.2}&13.3&15.1&16.6&13.5 \\ 
 \raisebox{1.5ex}[0pt]{$B_1$}&10&14.8&20.2&14.9&12.3&\textbf{26.2}&\textbf{27.7}&16.4&15.8&19.9&17.6 \\ 
\cline{2-2}\cline{3-3}\cline{4-6}\cline{7-8}\cline{9-10}\cline{11-12}
 &5&20.9&30.0&26.6&20.2&\textbf{40.4}&\textbf{40.1}&32.5&30.7&30.5&33.2 \\ 
 \raisebox{1.5ex}[0pt]{$B_2$}&10&31.4&52.5&51.1&35.1&\textbf{66.1}&\textbf{69.3}&51.3&52.9&51.6&52.2 \\ 
  \hline\hline 
 \end{tabular*}
\caption{Empirical power (in \%) for the F-test and multivariate permutation tests in a heteroskedastic 
three-sample setup ($\sigma_1=\sigma_2=1$, $\sigma_3=4$), bold font denotes the two most powerful permutation tests in each simulation setup.}\label{tabka1s}
\end{table}

The effect of heteroskedasticity is shortly investigated in a simulation summarized in Table~\ref{tabka1s}. The
setup remains exactly the same as in Table~\ref{tabka1}, but the standard deviation in the third group is set to $\sigma_3=4$. It is important to note that in this setup, with different standard deviations, the data are not exchangable even if $\mu_1=\mu_2=\mu_3$. Hence, the 
permutation test is actually testing the null hypothesis of `equality in distribution' even though the test 
statistics are based on sample mean differences. Thus, the rows denoted as $H_0$ in  Table~\ref{tabka1s} simply correspond to the setup where we take $\mu_1=\mu_2=\mu_3$ and one should not be surprised that the corresponding proportions of rejections exceed the nominal level $5 \%$.

Similarly as in the homoskedastic setup, the empirical power reported in Table~\ref{tabka1s}
is highest for the test based on optimal transport of the two-sided test statistics, followed by the remaining versions of the permutation test. Interestingly, the F-test actually performs quite poorly especially in the simulation setup $B_\delta$.

\begin{table}
\begin{tabular*}{\textwidth}{*{2}{c}@{\extracolsep{\fill}}r@{\extracolsep{\fill}}r@{\extracolsep{\fill}}r@{\extracolsep{\fill}}r@{\extracolsep{\fill}}r@{\extracolsep{\fill}}r@{\extracolsep{\fill}}r@{\extracolsep{\fill}}r@{\extracolsep{\fill}}r@{\extracolsep{\fill}}r} 
\hline\hline
& $(\mu_1,\mu_2,\mu_3,\mu_4)$ & 
$F$&$c(T)$&$c(F)$&$c(L)$&$t(\mathcal{P})$&$t(\mathcal{N})$&$t(\mathcal{P}^+)$&$t(\mathcal{N}^+)$&$p(\mathcal{P}^+)$&$p(\mathcal{N}^+)$ \\ 
 \cline{2-2} \cline{3-3} \cline{4-5} \cline{6-7} \cline{8-10} \cline{11-12} 
& $(0,0,0,0)$&4.3&5.0&5.2&5.6&4.4&5.2&5.0&4.0&4.9&4.6 \\ 
& $(0,0,3,1)$&{51.0}&\textbf{56.8}&41.6&32.2&45.5&35.4&\textbf{48.7}&42.6&37.4&30.4 \\ 
& $(0,0.5,2,1)$&{20.7}&\textbf{19.4}&\textbf{20.4}&17.0&16.9&13.1&{19.0}&{19.0}&{17.4}&15.1 \\ 
\raisebox{1.5ex}[0pt]{$N$} & $(0,1,2,3)$&{42.8}&38.9&\textbf{45.1}&41.0&36.5&27.1&\textbf{42.1}&38.1&{36.7}&33.5 \\ 
& $(0,2,-1,3)$&{78.0}&65.2&\textbf{75.4}&\textbf{73.5}&67.4&49.6&71.4&68.3&{62.1}&57.6 \\ 
& $(1,3,0,2)$&{45.7}&37.2&\textbf{46.1}&39.8&39.1&33.2&\textbf{44.6}&39.4&36.1&30.7 \\ 
 \cline{2-2} \cline{3-3} \cline{4-5} \cline{6-7} \cline{8-10} \cline{11-12} 
 &$(0,0,0,0)$&4.1&4.0&4.7&4.3&4.6&4.3&5.1&4.9&4.6&5.3 \\ 
 &$(0,0,3,1)$&{58.2}&\textbf{63.5}&54.4&32.4&52.8&44.8&\textbf{55.5}&52.2&40.9&31.2 \\ 
 &$(0,0.5,2,1)$&24.7&\textbf{29.5}&27.3&17.9&24.3&19.8&\textbf{27.4}&23.2&18.6&19.4 \\ 
  \raisebox{1.5ex}[0pt]{$\chi^2$}&$(0,1,2,3)$&{51.9}&43.7&\textbf{50.6}&48.2&\textbf{52.9}&33.5&49.2&47.4&41.7&41.0 \\ 
 &$(0,2,-1,3)$&77.7&66.8&\textbf{79.7}&77.1&75.5&47.0&\textbf{78.0}&74.4&66.3&63.5 \\ 
 &$(1,3,0,2)$&{51.0}&45.1&\textbf{53.9}&48.2&\textbf{50.5}&39.5&49.6&45.0&44.9&40.9 \\ 
 \hline 
 \hline 
 \end{tabular*}
\caption{Empirical power (in \%) in a four-sample problem, six scenarios, symmetric (centered Normal distribution) or skewed random errors (centered $\chi^2_2$ distribution) with variance 4, Helmert contrasts, $n_i=5$, $i\in\{1,\dots,4\}$, bold font denotes two most powerful permutation tests in each simulation setup.}\label{tabka-g4d3}\end{table}

\subsection{Four sample comparisons with Helmert contrasts}

Let us now consider a four sample problem, where  
$X_{j,i} = \mu_j + \varepsilon_{j,i}$, $j=1,\dots,4$, $i=1,\dots,n_j$ for $n_1=n_2=n_3=n_4$, with independent random errors generated either from a symmetric (centered normal distribution with variance $\sigma^2=4$) or skewed distribution (centered $\chi^2$ distribution with 2 degrees of freedom) with the same variance. The null hypothesis
\[
H_0:\mu_1=\mu_2=\mu_3=\mu_4
\]
will be tested using Helmert 
contrasts, that is via partial null hypotheses
\[
H_{0,1}:\mu_2-\mu_1=0,\quad H_{0,2}: \mu_3-\frac{\mu_2+\mu_1}{2} = 0, \quad H_{0,3}:\mu_4-\frac{\mu_1+\mu_2+\mu_3}{3}=0. 
\]
The corresponding test statistic $\bm{T}$ is three dimensional 
and its components are the  t-tests statistics corresponding to the three contrasts.
Again, the F-test of $H_0$ (based on $F$ quantiles) serves as a baseline  and 
it is compared to the various approaches to multivariate permutation tests. 


The simulation results summarized in Table~\ref{tabka-g4d3} show that the highest power is achieved either by 
the multivariate permutation tests with combining functions or by the multivariate permutation test using 
optimal transport of the partial permutation p-values to the positive part of the three-dimensional product form grid; note also that the optimal transport to the product form grid consistently outperforms the optimal 
transport to the three-dimensional non-product grid. 

This could  be partly explained by an insufficient grid size $B+1$, and it is expected that the difference in results for the two types of grids will vanish with increasing $B$. 
Recalculating the power for a two-sided test statistic with 
a non-product type grid in the last row of Table~4 using 200 simulations and a grid of size $B+1=8\,190$ leads to an empirical power 59 \% (instead of 39 \%), clearly outperforming all other tests in the same setup. Unfortunately, increasing the grid size is very computationally expensive.

It is also interesting to note that the power of all tests in  Table~4 is larger for the skewed random errors compared to the normal errors. This is, however, in agreement with some previous findings, see, e.g., \citet{tiku1971power}.


\begin{table}
\begin{tabular*}{\textwidth}{*{1}{c}@{\extracolsep{\fill}}c@{\extracolsep{\fill}}c@{\extracolsep{\fill}}c@{\extracolsep{\fill}}c@{\extracolsep{\fill}}c@{\extracolsep{\fill}}c} 
  \hline 
 \hline 
$(\mu_1,\mu_2,\mu_3,\mu_4)$ &$t(\mathcal{P})$&$t(\mathcal{N})$&$t(\mathcal{P}^+)$&$t(\mathcal{N}^+)$&$p(\mathcal{P}^+)$&$p(\mathcal{N}^+)$ \\ 
\cline{1-1} \cline{2-3} \cline{4-5} \cline{6-7} 
 $(0,0,3,1)$&0/100/0&0/100/0&8/84/8&8/84/8&11/79/10&10/80/10 \\ 
 $(0,0.5,2,1)$&5/93/2&6/93/1&18/65/17&19/64/16&19/63/18&20/64/15 \\ 
 $(0,1,2,3)$&10/32/58&\phantom{0}9/29/62&15/31/54&15/32/53&17/32/51&19/29/52 \\ 
 $(0,2,-1,3)$&20/27/53&20/30/51&20/25/55&21/28/51&25/29/46&23/30/47 \\ 
 $(1,3,0,2)$&43/51/6\phantom{0}&39/53/8\phantom{0}&39/48/13&38/48/14&37/47/16&38/47/15 \\ 
\hline 
 \hline 
 \end{tabular*}
\caption{Importance of marginal hypotheses (in \%) in a four-sample problem, five scenarios, Helmert contrasts, Normal distribution,   $\sigma=2$, $n_i=5$, $i\in\{1,\dots,4\}$.}\label{tabka-g4d3-pr}
\end{table}

 Table~\ref{tabka-g4d3-pr} summarizes the averaged  partial contributions $(D_1^2,D_2^2,D_3^2)\cdot 100 \% $ of the three underlying 
null hypothesis. Similarly as in Table~\ref{tabka1-prpi10}, it seems that the optimal transport approach allows 
to identify correctly the relative importance of the underlying marginal hypotheses.

\subsection{Two-sample comparisons of mean and variance} \label{sec:2mv2}

For the last Monte Carlo simulation 
consider a two sample problem, where observations from the $j$-th sample, $j\in\{1,2\}$, are generated 
as $X_{j,i} = \mu_j +\sigma_j \varepsilon_{j,i}$, $i=1,\dots,n$, where the independent random errors $\varepsilon_{j,i}$ have either symmetric (standard normal) or skewed (centered and standardized $\chi^2_2$ distribution). In both simulation setups $EX_{j,i}=\mu_j$ and $\Var X_{j,i}=\sigma_j^2$ and we are interested in testing the
composite null hypothesis $H_0: \mu_1=\mu_2 \ \& \ \sigma_1^2=\sigma_2^2$ against general alternatives. This is a special case of the problem described in Section~\ref{sec:2.3.1} for $K=2$ samples and $r=2$ moments. 

Obviously,
the means can be compared using a two-sample t-test statistic, say $T_{12}$, and the ratio of sample variances
$F_{12}=S^2_1/S^2_2$ can be used for testing equality of variances. Apart of performing these two tests
separately,
using the critical values based on $t$- and $F$-distribution, we apply also the multivariate permutation 
tests utilizing the bivariate test statistic $(T_{12},\log(F_{12}))^\top$, where $\log(F_{12})$ is used mainly
in  order to facilitate the calculations for the two- and one-sided tests used in the optimal
transport approach.

\begin{table}
\begin{tabular*}{\textwidth}
{*{2}{c}@{\extracolsep{\fill}}r@{\extracolsep{\fill}}r@{\extracolsep{\fill}}r@{\extracolsep{\fill}}r@{\extracolsep{\fill}}r@{\extracolsep{\fill}}r@{\extracolsep{\fill}}r@{\extracolsep{\fill}}r@{\extracolsep{\fill}}r@{\extracolsep{\fill}}r@{\extracolsep{\fill}}r@{\extracolsep{\fill}}r} 
\hline\hline
&$\mu_2$ & $\sigma_2$ & $T_{12}$ & $F_{12}$ &$c(T)$&$c(F)$&$c(L)$&$t(\mathcal{P})$&$t(\mathcal{N})$&$t(\mathcal{P}^+)$&$t(\mathcal{N}^+)$&$p(\mathcal{P}^+)$&$p(\mathcal{N}^+)$ \\ 
 \cline{2-3}\cline{4-14}
 &0 & 2 &5.3&5.4&5.5&5.9&5.4&5.3&4.6&4.8&4.6&5.5&5.8 \\ 
 &2 & 2 &{59.1}&6.3&\textbf{45.3}&43.1&35.4&39.6&35.8&40.8&\textbf{43.6}&41.5&43.3 \\ 
 \raisebox{1.5ex}[0pt]{$N$} &0 & 4&3.9&{48.7}&\textbf{30.2}&\textbf{32.7}&23.6&25.9&23.4&29.8&28.6&25.9&26.9 \\ 
 &2&4 &26.1&47.3&43.3&\textbf{50.8}&\textbf{49.8}&40.8&34.1&47.6&49.4&42.9&47.3 \\ 
 \cline{2-3}\cline{4-14} 
& 0&2&4.9&23.0&5.1&4.5&5.2&5.1&4.5&4.6&5.1&5.0&5.7 \\ 
& 2&2&60.9&23.4&62.9&52.4&39.9&\textbf{77.7}&   \textbf{73.9}&62.7&57.7&64.2&60.0 \\ 
\raisebox{1.5ex}[0pt]{$\chi_2^2$} &0&4&7.5&50.1&29.4&22.6&15.8&\textbf{39.4}&\textbf{39.8}&30.2&29.8&32.7&32.0 \\ 
& 2&4&24.2&52.1&\textbf{24.9}&\textbf{25.4}&24.5&17.1&15.2&20.5&21.7&22.5&21.2 \\ 
\hline\hline 
 \end{tabular*}
\caption{Empirical level and power (in \%) in a two-sample problem combining the tests of equality of
expectations ($T_{12}$, two-sample t-test) and variances ($F_{12}$, F test for equality of variance), $\mu_1=0$, $\mu_2\in\{0,2\}$, $\sigma_1=2$,
$\sigma_2\in\{2,4\}$, $n_1=n_2=10$, level $\alpha=0.05$, 1000 simulations, standard normal or standardized $\chi^2_2$ random errors, bold font denotes two most powerful permutation tests in each simulation setup.}\label{tabka-2mv2}
\end{table}

Table~\ref{tabka-2mv2} reports the percentage of rejection of $H_0$ for several 
 tests under various alternatives. 
For normally distributed random errors, all tests behave as expected, with overall best behaviour observed for the Fisher combining function.
Looking at the optimal transport-based tests, the power seems to be more
stable (with respect to the grid choice) and only slightly
smaller than the power achieved by the best combining function.

In the second part of Table~\ref{tabka-2mv2}, corresponding to non-symmetrically
distributed random errors, 
one can observe a
slightly larger power of the two-sample t-test and a complete 
failure of the variance
equality F-test (with an empirical level equal to 23\%)
 that is in agreement with its well-known 
non-robustness  \citep[Section 11.3]{lehman}. Concerning the
permutation tests, two-sided test statistics clearly outperform
all other tests if only one of the marginal hypotheses does not
hold. One-sided approaches using either the one-sided test
statistics or p-values lead similar results in most setups. Hence, one can conclude that the
optimal transport approach seems to be more stable, with slightly
better performance in setups with only one violated marginal
hypothesis.

 \begin{table}
\begin{tabular*}{\textwidth}{*{3}{c}@{\extracolsep{\fill}}c@{\extracolsep{\fill}}c@{\extracolsep{\fill}}c@{\extracolsep{\fill}}c@{\extracolsep{\fill}}c@{\extracolsep{\fill}}c} 
 \hline
 \hline
&$\mu_2$ & $\sigma_2$ & $t(\mathcal{P})$&$t(\mathcal{N})$&$t(\mathcal{P}^+)$&$t(\mathcal{N}^+)$&$p(\mathcal{P}^+)$&$p(\mathcal{N}^+)$ \\ 
 \cline{2-3}\cline{4-9}
& 2&2&100/0\phantom{00}&100/0\phantom{00}&89/11&88/12&86/14&85/15 \\ 
$N$& 0&4&\phantom{00}0/100&\phantom{00}0/100&16/84&17/83&18/82&19/81 \\ 
& 2&4&38/62&40/60&42/58&43/57&39/61&41/59 \\ 
 \cline{2-3}\cline{4-9}
 &2&2&88/12&90/10&91/9&89/11&85/15&84/16 \\ 
$\chi^2_2$ &0&4&20/80&18/82&19/81&17/83&17/83&19/81 \\ 
 &2&4&78/22&77/23&66/34&65/35&63/37&62/38 \\  
 \hline 
 \hline
 \end{tabular*}
 \caption{Importance of partial tests (in \%) in a two-sample problem combining the tests of equality of
expectations and variances.}\label{tabka-2mv2-pr} 
\end{table}

The percentages of contributions  of partial tests are summarized in
Table~\ref{tabka-2mv2-pr}. The results for the normal distribution
seem to be more readable, with the percentages quite clearly identifying
the correct alternatives, especially in the
two-sided approach. For skewed random errors, the empirical contributions seem to be
a bit more ``fuzzy'' but, in all cases, the percentages corresponding
to valid null hypotheses are smaller than 20\% whereas the
percentages corresponding to valid alternatives are always
greater than 80\%.

\section{Application to functional data}\label{sec:fanova}

\begin{figure}
    \includegraphics[width=10cm]{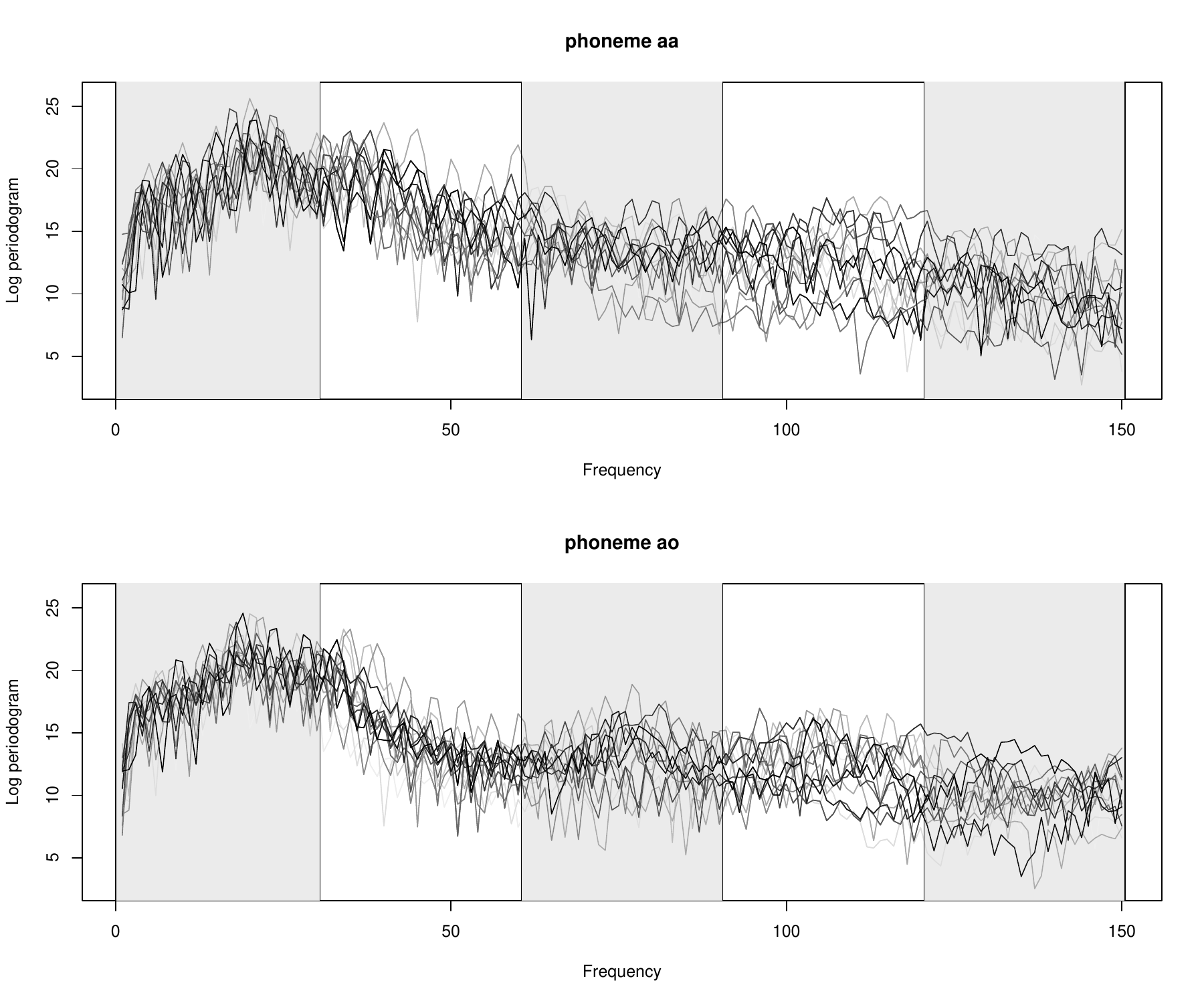}
\caption{Log periodograms for phonemes ``aa'' and ``ao'', $n_1=n_2=15$,
      with frequencies split to five segments, source: \textsf{R} library \texttt{fds} \citep{fds}.
      }\label{fig-phoneme}
\end{figure}

\citet{hlavka2022functional} investigated the phoneme data set plotted in
Figure~\ref{fig-phoneme} and observed statistically significant differences in the
distribution of the two functional random samples (using a test comparing empirical characteristic functions) 
in the first three of five segments (with sample size 15 in each sample). These results were confirmed by applying separate permutation tests 
of equality of mean functions (test statistic Fmax) \citep{gorecki19} and covariance operators (test statistic SQ based on square-root distance) \citep{cabassi2017permutation}
reported in the first two rows in Table~\ref{tab-phoneme}.

\begin{table}
\begin{tabular*}{\textwidth}{l*{5}{@{\extracolsep{\fill}}c}} 
\hline
\hline
&       1. &  2.&   3. &  4. & 5.  \\  
\cline{2-2}\cline{3-3}\cline{4-4}\cline{5-5}\cline{6-6}
\cline{1-1}\cline{2-2}\cline{3-3}\cline{4-4}\cline{5-5}\cline{6-6}
mean (Fmax)       & 0.443& {0.002}& {0.001}& 0.828&0.708 \\
covariance (SQ)   & {0.039}& {0.005}& 0.075& 0.505& 0.123 \\
\cline{1-1}\cline{2-2}\cline{3-3}\cline{4-4}\cline{5-5}\cline{6-6}
Tippett           & 0.071 & {0.003} & {0.001} & 0.752 & 0.227 \\
Fisher           & 0.113 & {0.001} & {0.001} & 0.775 & 0.287 \\
Liptak           & 0.138 & {0.001} & {0.001} & 0.751 & 0.329 \\
\cline{1-1}\cline{2-2}\cline{3-3}\cline{4-4}\cline{5-5}\cline{6-6}
  $t(\mathcal{P}^+)$                & {0.048 (8\%)} & {0.048 (95\%)} &{0.048 (100\%)} &0.762  &  0.286 \\
  $t(\mathcal{N}^+)$                & {0.081 (4\%)} & {0.011 (92\%)} & {0.001 (100\%)} & 0.797 & 0.274 \\
\cline{1-1}\cline{2-2}\cline{3-3}\cline{4-4}\cline{5-5}\cline{6-6}
  $p(\mathcal{P}^+)$               & {0.095 (4\%)}  & {0.048 (52\%)} & {0.048 (61\%)} & 0.762  & 0.286 \\            
  $p(\mathcal{N}^+)$            & {0.059 (6\%)}    & {0.018 (51\%)} & {0.013 (63\%)} & 0.762 & 0.295 \\
\hline
\hline
\end{tabular*}
\caption{P-values of permutation tests for equality of mean functions (Fmax) and covariance operators (SQ) in five segments of the phoneme data set together with p-values obtained by nonparametric combination and optimal transport, $n_1=n_2=15$, importance of the first marginal hypothesis (in $\%$) is calculated for optimal transport based permutation tests.}
\label{tab-phoneme}
\end{table}

As described in Section~\ref{sec:func}, the joint null hypothesis of equality of the mean functions and the covariance operator can be tested using the multivariate permutation test. In Table~\ref{tab-phoneme}, we report the p-values obtained by 
nonparametric combination (Tippett, Fisher, and Liptak combining functions) and also p-values obtained by the optimal transport of the one-sided test statistic or partial permutation p-values. In segment~1, the composite null hypothesis is rejected only using the optimal transport approach of the test statistic on the product form grid $\mathcal{P}^+$ although p-values obtained by other approaches are also quite close to 0.05. The percentages (8\%/92\%) indicate that the significant difference between the two samples in the first segment is caused mostly by the violation of the second null hypothesis and the reason for rejecting the composite null hypothesis in the first segment is a different covariance structure. 

In the remaining segments, all multivariate permutation tests arrive to the same conclusions. The p-value based on the product form grid sometimes seems to be somewhat larger but $0.048\doteq 1/21$ is actually the smallest possible p-value (corresponding to the outermost ring in the product form grid). The partial contributions obtained for one-sided test statistic suggest that the differences in mean functions are more responsible for rejecting the null hypothesis in segments~2 and~3. Interestingly, the partial contributions obtained for the p-values seem to have a different interpretation since, in this case, both partial p-values are quite small and, in this sense, i.e., from a point of view of the partial p-values, both marginal null hypotheses contribute almost equally to the rejection of the composite null hypothesis.

\section{Summary}\label{sec:5}

This paper proposes to combine the optimal transport methods with multivariate permutation tests.
It is shown that the empirical transport of a multivariate permutation
test statistics leads to a well-defined single p-value, which can be also interpreted as a measure of conformity of the observed multivariate test statistic with 
the (simulated) permutation sample.

In contrast to the classical approach to multivariate permutation tests
utilizing the so-called nonparametric combination, the 
proposed method allows to evaluate also the 
contributions of the partial test statistics to the rejection of
the null hypothesis, providing additional insights. 
The definition of absolute contributions guarantees that the null hypothesis is rejected if and only if the sum of these absolute contributions is larger than the desired confidence level. 
For a large number of comparisons, these relative
contributions could be presented in a suitable graph, similar to the scree plot of eigenvalues in principal component analysis, summarizing the influence of partial test statistics and
indicating main reasons for rejecting the composite null hypothesis.

%


The Monte Carlo simulation study reveals that, in some setups, the optimal transport approach to
multivariate permutation tests clearly outperforms the classical
nonparametric combination. This holds especially in setups, in
which the transformation to p-values (or
one-sided test statistics) leads to some loss of information (see
Table~\ref{tabka1s}), with skewed random errors, or when some of the marginal
null hypotheses hold true  (see Table~\ref{tabka-2mv2}). 
In other situations, nonparametric combination with properly chosen combining
function performs very well 
but, unfortunately, the best combining function usually cannot be reliably chosen in advance unless we have some prior information concerning the alternative. Therefore, in general, it may be safer to use the optimal transport approach, circumventing the possibility of choosing the combining function inappropriately.

In a simulation study, we have compared two asymptotically equivalent grids
that typically lead comparable results in lower-dimensional problems, although
the behavior of the product-type grid seems to be a bit more stable. The simulation study also suggests that the number of points should be larger for two-sided test statistics and for the non-product type grid.
Theoretically, the proposed approach is directly applicable for
an arbitrary number of comparisons but, since the grid size should
increase exponentially with the
dimension, the necessary calculations may soon become computationally  unfeasible. Speeding up these calculations thus remains a challenging future research problem that 
may make the optimal transport approach applicable even with large number of simultaneous comparisons.


\bibliographystyle{apalike}
\bibliography{short,glp-body}

\end{document}